\documentclass[twocolumn,secnumarabic,amssymb, nobibnotes, aps, prd]{revtex4-2}

\usepackage{dcolumn}% Align table co
\usepackage{bm} % bold math
\usepackage{graphicx} % Include figure files
\usepackage{color}
\usepackage{subfigure}
\usepackage{hyperref}
\usepackage{latexsym}
\usepackage{amsthm}
\usepackage{amssymb}
\usepackage{verbatim}
\usepackage{wasysym}

\DeclareGraphicsExtensions{.jpg, .pdf, .mps, .png, .eps, .ps, .EPS, .gif}
\begin{document}
\def\be{\begin{equation}}
\def\ee{\end{equation}}

\def\bc{\begin{center}}
\def\ec{\end{center}}
\def\bea{\begin{eqnarray}}
\def\eea{\end{eqnarray}}

\newcommand{\avg}[1]{\langle{#1}\rangle}
\newcommand{\Avg}[1]{\left\langle{#1}\right\rangle}

\newcommand{\cch}[1]{\left[#1\right]}
\newcommand{\chv}[1]{\left \{ #1\right \} }
\newcommand{\prt}[1]{\left(#1\right)}
\newcommand{\aver}[1]{\left\langle #1 \right\rangle}
\newcommand{\abs}[1]{\left| #1 \right|}

\def\ie{\textit{i.e.}}
\def\etal{\textit{et al.}}
\def\m{\vec{m}}
\def\G{\mathcal{G}}
\def\fig{FIG.}
\def\tab{TABLE }

\newcommand{\gin}[1]{{\bf\color{magenta}#1}}
\newcommand{\bob}[1]{{\bf\color{red}#1}}
\newcommand{\bobz}[1]{{\bf\color{magenta}#1}}

\title{Universal behavior of site and bond percolation thresholds on regular lattices with compact extended-range neighborhoods in 2 and 3 dimensions}

\author{Zhipeng Xun}
\email{zpxun@cumt.edu.cn}
\author{Dapeng Hao}
\email{dphao@cumt.edu.cn}
\affiliation{School of Material Sciences and Physics, China University of Mining and Technology, Xuzhou 221116, China}
\author{Robert M. Ziff}
\email{rziff@umich.edu}
\affiliation{Center for the Study of Complex System and  Department of Chemical Engineering, University of Michigan, Ann Arbor, Michigan 48109-2800, USA}

\date{\today}

\begin{abstract}
Extended-range percolation on various regular lattices, including all eleven Archimedean lattices in two dimensions, and the simple cubic (\textsc{sc}), body-centered cubic (\textsc{bcc}), and face-centered cubic (\textsc{fcc}) lattices in three dimensions, is investigated.  In two dimensions, correlations between coordination number $z$ and site  thresholds $p_c$ for Archimedean lattices up to 10th nearest neighbors (NN) are seen by plotting $z$ versus $1/p_{c}$ and $z$ versus $-1/\ln(1-p_c)$, using the data of d'Iribarne et al.\ [J. Phys. A 32:2611, 1999] and others. The results show that all the plots overlap on a line with a slope consistent with the theoretically predicted asymptotic value of $zp_{c} \sim 4 \eta_c = 4.51235$, where $\eta_c$ is the continuum threshold for disks. In three dimensions, precise site and bond thresholds for \textsc{bcc} and \textsc{fcc} lattices with 2nd and 3rd NN, and bond thresholds for the \textsc{sc} lattice with up to the 13th NN, are obtained by Monte-Carlo simulations, using an efficient single-cluster growth method. For site percolation, the values of thresholds for different types of lattices with compact neighborhoods also collapse together, and linear fitting is consistent with the predicted value of $zp_{c} \sim 8 \eta_c =  2.7351$, where $\eta_c$ is the continuum threshold for spheres. For bond percolation, Bethe-lattice behavior $p_c = 1/(z-1)$ is expected to hold for large $z$, and the finite-$z$ correction is confirmed to satisfy $zp_{c} - 1 \sim a_{1}z^{-x}$, with $x=2/3$ for three dimensions as predicted by Frei and Perkins [Electron.\ J. Probab.\ 21:56, 2016] and by Xu et al.\ [Phys.\ Rev.\ E, 103:022127, 2021]. Our analysis indicates that for compact neighborhoods, the asymptotic behavior of $zp_{c}$ is universal, depending only upon the dimension of the system and whether site or bond percolation, but not upon the type of lattice.
\end{abstract}


\maketitle
\section{Introduction}
It is well known that percolation is an important model in statistical physics \cite{BroadbentHammersley1957,StaufferAharony1994}. As a paradigmatic model, it can describe diverse phenomena in various fields, such as liquids moving in porous media \cite{BolandtabaSkauge2011,MourzenkoThovertAdler2011}, forest-fire problems \cite{Henley1993,GuisoniLoscarAlbano2011} and epidemics \cite{MooreNewman2000,Ziff2021}. Considering percolation on a lattice, each edge (vertex) is occupied by a bond (site) with probability $p$, and clusters of neighboring occupied sites or connected bonds can be constructed. As $p$ increases, the clusters become larger, and at the threshold point $p_{c}$, an infinite cluster spanning over the whole lattice emerges. Over the last several decades, a tremendous amount of work has gone into finding exact or approximate values of the percolation thresholds for a variety of systems, as well as finding formulas to approximately predict those thresholds.

Many kinds of percolation models have been established. The most common one is to occupy sites or bonds on a regular lattice with statistically independent probability $p$.  Site and bond percolation can be distinguished depending on the method of obtaining the cluster. One can also consider continuum percolation systems \cite{XuWangHuDeng2021,MertensMoore2012,QuintanillaZiff2007,TarasevichEserkepov20}, such as overlapping disks and spheres placed randomly. Further variations involve correlated percolation \cite{Kantor1986,ZierenbergFrickeMarenzSpitznerBlavatskaJanke2017}, like for example drilling percolation \cite{SchrenkHilarioSidoraviciusAraujoHerrmannThielmannTeixeira2016,Grassberger2017}. In bootstrap percolation \cite{ChoiYu2020,ChoiYu2019,MuroBuldyrevBraunstein2020,MuroValdezStanleyBuldyrevBraunstein2019}, sites and/or bonds are first occupied and then successively culled from a system if a site does not have at least $k$ neighbors. Another important model of percolation, which is in a different universality class, is directed percolation \cite{WangZhouLiuGaroniDeng13,Grassberger2009,Grassberger2009b,Jensen2004}, where connectivity along a bond depends upon the direction of the flow. 
%For more information, one can search ``percolation", ``percolation theory", ``percolation threshold", ``percolation critical exponents", et al, on Wikipedia. 

Among numerous models, percolation on lattices with extended neighborhoods has been of longstanding interest. In fact, this kind of percolation system has many applications. For example, site percolation on lattices with extended neighborhoods relates to problems of adsorption of extended shapes on a lattice, such as disks and squares \cite{KozaKondratSuszcaynski2014,KozaPola2016}, and bond percolation with extended neighbors has long-range links similar to small-world networks \cite{Kleinberg2000}. In addition, bond percolation with extended neighbors is also similar to spatial models of the spread of epidemics via long-range links \cite{SanderWarrenSokolov2003}. Just recently \cite{Ziff2021}, it was pointed out how the threshold is related to the basic epidemic infectivity parameter $R_{0}$, for trees (Bethe lattice), trees with triangular cliques, and non-planar lattices with extended-range connectivity.

The investigation of percolation on lattices with extended neighborhoods dates back to the ``equivalent neighbor model" of Dalton, Domb and Sykes from 1964 \cite{DaltonDombSykes64,DombDalton1966,Domb72}, and numerous studies have appeared since then. Extended-range site percolation on compact regions in a diamond shape on a square lattice, up to lattice distance of 10, was studied by Gouker and Family \cite{GoukerFamily83}. Other lattices, including \textsc{bcc} and \textsc{fcc} with extended neighborhoods, have also been studied \cite{JerauldScrivenDavis1984,GawronCieplak91}. d'Iribarne, Rasigni and Rasigni \cite{dIribarneRasigniRasigni95,dIribarneRasigniRasigni99,dIribarneRasigniRasigni99b} studied site percolation on all eleven Archimedean lattices (``mosaics") with extended-range connections up to the 10th nearest neighbors (NN), which we will discuss in detail later.  It has been suggested that these results may be applicable to a model of constrained percolation \cite{ReimannTupak02}.
Malarz and coworkers \cite{MalarzGalam05,GalamMalarz05,MajewskiMalarz2007,KurzawskiMalarz2012,Malarz2015,KotwicaGronekMalarz19,Malarz2020,Malarz21} carried out extensive numerical simulations on lattices with combinations of ``complex neighborhoods" in two, three, and four dimensions. Koza and collaborators \cite{KozaKondratSuszcaynski2014,KozaPola2016} studied percolation of overlapping shapes on a lattice, which can be mapped to extended-range site percolation.  While much of the earlier work concerned site percolation, bond percolation on extended lattices has been studied more extensively recently \cite{OuyangDengBlote2018,DengOuyangBlote2019,XunZiff2020,XunZiff2020b,XuWangHuDeng2021}.  

Many studies have focused on exploring the correlations between percolation thresholds $p_{c}$ and coordination number $z$ or other properties of lattices. It has been argued \cite{Domb72,dIribarneRasigniRasigni99b,KozaKondratSuszcaynski2014,KozaPola2016} that for extended-range site percolation, the threshold $p_c$ for large $z$ can be related to the continuum percolation threshold $\eta_c$ for objects of the same shape as the neighborhood, and this relationship is further clarified \cite{XunHaoZiff2021} by $p_c \sim 2^{d} \eta_{c}/z$, where $d$ is the dimension of the system. In this paper, we show that the asymptotic behavior of $z p_c$ for site percolation,  
\begin{equation}zp_c \sim 2^{d} \eta_{c}
\label{eq:zpcsite}
\end{equation}
is indeed universal, for neighborhoods limiting to a circle or sphere for $z$ large. In the first task of this paper, we investigate site percolation of the 11 two-dimensional Archimedean lattices and the three-dimensional \textsc{sc}, \textsc{bcc}, \textsc{fcc} lattices with extended neighbor connections. By fitting the data to the forms $z$ vs $1/p_c$ and $z$ versus $-1/\ln(1-p_c)$, we find good agreement with the predicted behavior of Eq.\ (\ref{eq:zpcsite}) with $4 \eta_c = 4.51235$ for lattices in two dimensions, and $8 \eta_c =  2.7351$ for lattices in three dimensions, suggesting a universal asymptotic behavior of $zp_c$ for lattices with compact extended neighborhoods.

For bond percolation, one expects that Bethe-lattice behavior $p_c = 1/(z-1)$ to hold for large $z$, because for large $z$ and small $p$, the chance of hitting the same site twice is low and the system behaves basically like a tree. Theoretical analysis of finite-$z$ corrections for bond thresholds has recently been given by Frei and Perkins \cite{FreiPerkins2016} and Xu et al.\ \cite{XuWangHuDeng2021} as 
\begin{equation}
zp_{c} - 1 \sim a_{1}z^{-x}
\label{eq:zpcbond}
\end{equation}
where $x = (d-1)/d$ for $d = 2, 3$, implying $x = 1/2 $ in two dimensions and $2/3$ in three dimensions. In the second task of this paper, we study bond percolation on three-dimensional \textsc{sc}, \textsc{bcc}, and \textsc{fcc} lattices with extended neighborhoods by Monte-Carlo simulation using a single-cluster growth algorithm. We find many precise bond percolation thresholds, and data fitting is consistent with Eq.\ (\ref{eq:zpcbond}) with $x=2/3$.

The remainder of the paper is organized as follows. Section \ref{sec:theory} describes the theoretical prediction of the asymptotic behavior of $zp_{c}$ for site percolation. The results and analysis in two and three dimensions are given in Sec.\ \ref{sec:archimedean} and Sec.\ \ref{sec:3d}, respectively, and in Sec.\ \ref{sec:conclusions} we present our conclusions. 

\section{Theoretical analysis}
\label{sec:theory}

We analyze the effective extended-range neighborhood for an object of an arbitrary shape.
%The relation between continuum percolation and lattice percolation for neighborhoods of a general shape can be generalized to find a formula for large $z$ behavior of $p_c$.
First we consider a continuum system of volume $V$ with the random placement of $N$ overlapping objects of the given shape. The continuum percolation threshold $\eta_c$ represents the total volume fraction of the adsorbed objects, including overlapping volume, at the critical point
\begin{equation}
    \eta_c= a_d r^d \frac{N}{V}, 
\label{eq:sphere}
\end{equation}
where $r$ is the radius or other length scale of the object and $a_d r^d$ is its volume, with $a_d$ depending upon the shape of the object. Covering the space occupied by the objects with a fine mesh of any lattice type, the system can be mapped to site percolation on that lattice with an extended neighborhood of essentially the same shape but with a length scale $2r$ about the central point. The ratio $N v_0 /V = p_c$ corresponds to the site occupation threshold on the lattice, where $v_0$ is the area or volume per site.  The effective $z$ is equal to the number of sites within that region of influence of length scale $2r$
\begin{equation}
    z = a_d (2r)^d/v_0
    \label{eq:zad}
    \end{equation}
For example, for squares or cubes of length $2k$ on a square or cubic lattice with $v_0 = 1$, \cite{Koza19}, $z = (2k)^d$.  Then it follows from Eqs.\ (\ref{eq:sphere}) and (\ref{eq:zad}) that  
\begin{equation}
    zp_c = (a_d (2r)^d/v_0) (N v_0 /V) = 2^d (a_d r^d N/V) = 2^d \eta_c.
\label{eq:zpc}
\end{equation}
as given in Eq.\ (\ref{eq:zpcsite}).  This equation should describe the behavior of $p_c$ for large $z$ where the objects become similar to a continuum, for systems with compact neighborhoods, and where $\eta_c$ is the critical coverage for continuum systems of the shape of the neighborhood.

\begin{figure}[htbp] % figure placement: here, top, bottom, or page
\centering
\includegraphics[width=3.2in]{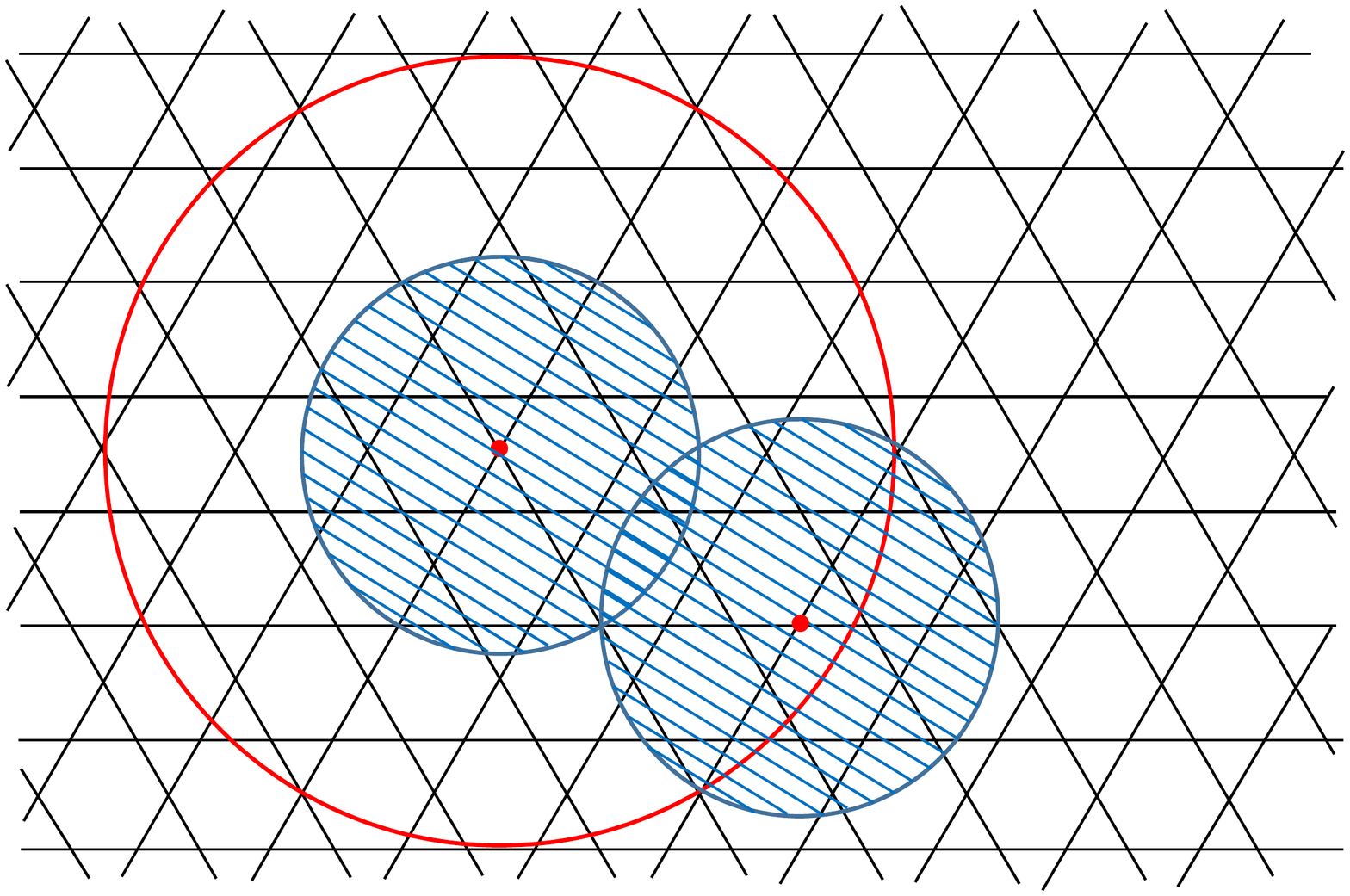}
\caption{Illustration showing the relation between continuum percolation of disks of radius 3, and the equivalent extended-range site percolation model of range 6, embedded on a kagome lattice.  The smaller disc covers 27 sites while the larger one covers 97 sites, and is equivalent to having $z=96$ (we do not count the central site when considering $z$) with a range up to the 15th NN.}
\label{fig:extendedrange}
\end{figure}

Figure \ref{fig:extendedrange} illustrates the situation where we have the continuum percolation of disks, here of radius 3, embedded on a kagome lattice of unit edge length.  The system is seen to be equivalent to an extended-range percolation model with the centers of the disks being sites connected if they fall within a radius 6 with each other.  The disk of radius 3 covers 27 sites on the lattice, while the range of the equivalent site percolation model covers 97 sites and extends to the 15th NN.  Note here the number of sites in circle of radius 6 (97 sites) is not four times the number as in the disk (27 sites), as that ratio would be for larger circles or indeed objects of any shape according to Eq.\ (\ref{eq:zad}).

We will choose neighborhoods with sites that fall within a radius $r$.  As $z \rightarrow \infty$, the shape of the neighborhood becomes more circular (or spherical), so one can naturally suppose that the asymptotic behavior of $zp_c$ should be universal, with $\eta_c$ for a disk or spheres depending upon the dimension of the system. For circular neighborhoods in two dimensions, where $\eta_c$ for disks equals 1.128087 \cite{XuWangHuDeng2021,MertensMoore2012,QuintanillaZiff2007,TarasevichEserkepov20,LiOstling16}, one should thus expect
\begin{equation}
    zp_c \sim 4.51235,
\label{eq:approx2d}
\end{equation}
for large $z$, while for spherical neighborhoods in three dimensions, where $\eta_c$ for spheres equals 0.34189 \cite{LorenzZiff2000,TorquatoJiao2012,GoriTrombettoni15}, one should expect
\begin{equation}
    zp_c \sim 2.7351.
\label{eq:approx3d}
\end{equation}

A related approach comes frem Ref.\ \cite{KozaKondratSuszcaynski2014}, where Koza et al.\ investigated squares of size $k \times k$ and cubes of size $k \times k \times k$, randomly distributed in an overlapping manner on square or cubic lattices. The critical number (per site) of these objects for percolation between neighboring occupied sites defines the threshold $p_c$. At the percolation point, the fraction of sites on the lattice covered by at least one square or cube, $\phi_{c} (k)$, can be related to $p_c$ by $\phi_c(k) = 1 - (1-p_c)^{k^d}$, which can also be written as
\begin{equation}
    p_c = 1 - (1 - \phi_c(k))^{1/k^d}. 
\label{eq:phick}
\end{equation}
For large $k$, the model limits to the  percolation of aligned squares or cubes on a continuum, and $\phi_c(k)$ limits to $\phi_c$ for the corresponding continuum system. Replacing $\phi_c(k)$ by the continuum value $\phi_c$ in Eq.\ (\ref{eq:phick}), one obtains an approximation to find $p_c$ for discrete $k^d$ objects with large but finite $k$ \cite{KozaKondratSuszcaynski2014,KozaPola2016} 
\begin{equation}
    p_c = 1 - (1 - \phi_c)^{1/k^d}. 
\label{eq:phic}
\end{equation}

One defines $\eta_c$ as the total area or volume of the objects placed or adsorbed in the system, including the area or volume of the overlapped parts, divided by the area or volume of the system.  The quantity $\eta_c$ is related to $\phi_c$ by $\phi_c = 1-e^{-\eta_c}$
which can be substituted into Eq.\ (\ref{eq:phic}) to yield 
\begin{equation}
    p_c = 1-e^{-\eta_c/k^d}
    \label{eq:pcetac}
\end{equation}

We can  generalize Eq.\ (\ref{eq:pcetac}) to objects of arbitrary shape and an arbitrary lattice, by replacing $k^d$ by $z/(2^d)$ (the number of sites in the neighborhood), yielding the general formula
\begin{equation}
    p_{c} = 1 - e^{-2^d \eta_c/z}
\label{eq:general}
\end{equation}
The above formula does not depend upon the type of lattice (square, triangular, etc.) used, because for lattices with $z$ NN, the number of sites of the equivalent object (disk, sphere, etc.) is always $z/(2^d)$.  The type of lattice does not matter because the volume per lattice site $v_0$ cancels out, as seen in Eq.\ (\ref{eq:zpc}).  Note that Eq.\ (\ref{eq:general}) was also given in Ref.\ \cite{XunHaoZiff2021}, although without a complete derivation as given here.

Solving for $z$, Eq.\ (\ref{eq:general}) yields
\begin{equation}
    z = \frac{2^d \eta_c}{-\ln(1-p_c)}
\label{eq:lnpc}
\end{equation}

In the limit of large $z$, Eq.\ (\ref{eq:general}) limits to Eq.\ (\ref{eq:zpcsite}) and this gives an alternative derivation of that result.  However, for moderate $z$, it has been found \cite{XunHaoZiff2021} that for some systems, Eq.\ (\ref{eq:general}) gives a better estimate of $p_c$ than Eq.\ (\ref{eq:zpcsite}).  One of the goals of this paper is to compare Eqs.\ (\ref{eq:general}) and (\ref{eq:zpcsite}) in modeling the finite-$z$ behavior.

In Ref.\ \cite{XunHaoZiff2021}, we also found that the finite-$z$ effect can be taken into account by assuming $p_c = c/(z+b)$, where $b$ and $c=2^d \eta_c$ are constants. We can write this relation as 
\begin{equation}
    z = \frac{c}{p_c} - b.
\label{eq:z}
\end{equation}
In contrast to Eq.\ (\ref{eq:lnpc}), this formula contains a new adjustable parameter, $b$.
For more details about these formulas, one can also see Refs.\ \cite{KozaKondratSuszcaynski2014,KozaPola2016,XunHaoZiff2021}.

Equations (\ref{eq:lnpc}) and (\ref{eq:z}) show that if we plot $z$ versus $-1/\ln(1-p_{c})$ or $z$ 
versus $1/p_{c}$, one can directly get the value of $c = 2^{d} \eta_c$ from the slopes, and in the latter case, the value of $-b$ from the intercept.

\begin{figure}[htbp] %  figure placement: here, top, bottom, or page
\centering
\includegraphics[width=3.0in]{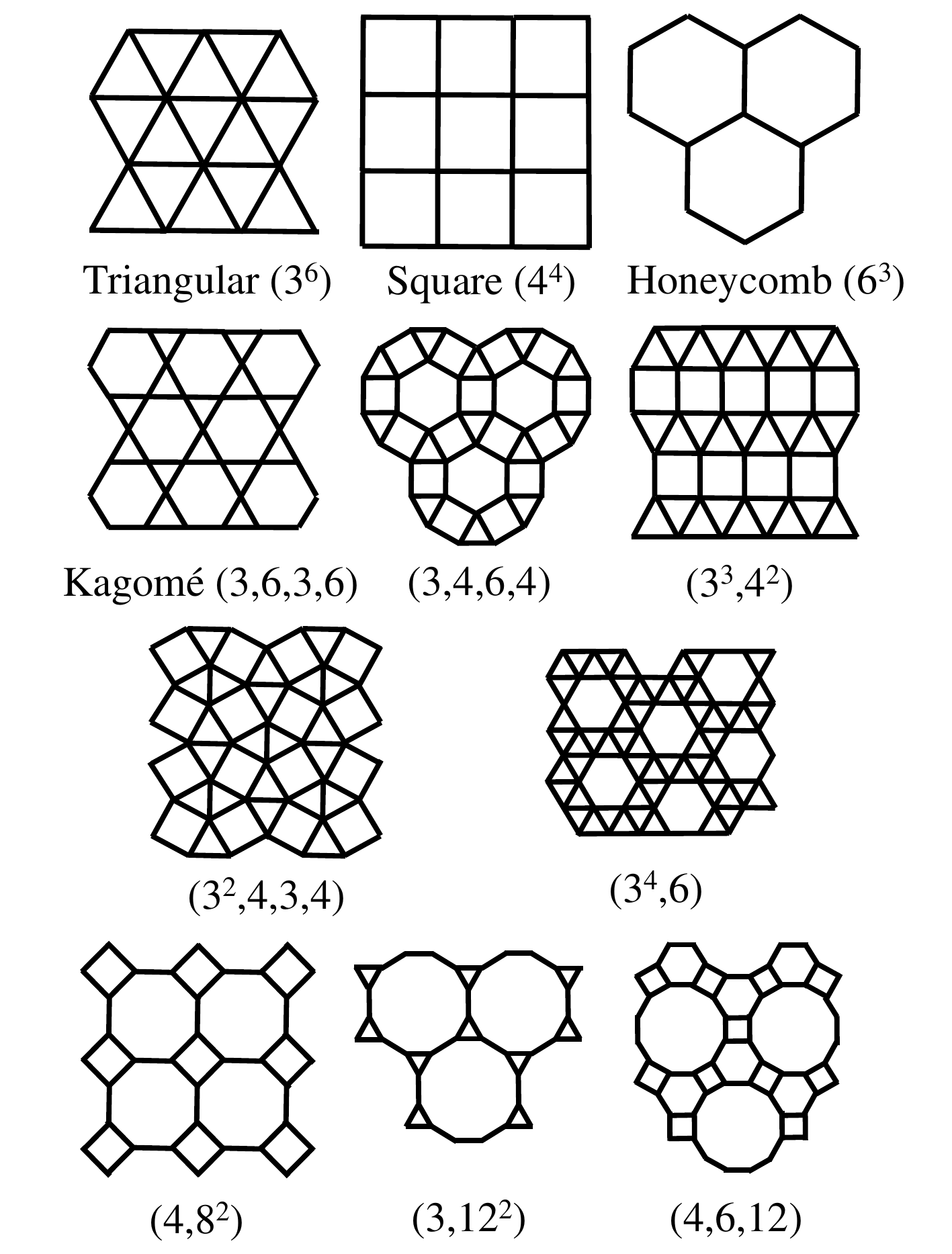}
\caption{Illustrations of the 11 Archimedean lattices.}
\label{fig:archimedean}
\end{figure}

\section{Site percolation on Archimedean lattices with extended connections}
\label{sec:archimedean}
Figure \ref{fig:archimedean} shows drawings of the 11 Archimedean lattices, in which all polygons are regular and each vertex is surrounded by the same sequence of polygons.  Each lattice is characterized by a standard notation; for example, the notation $(3^4,6)$ means that each vertex is surrounded by four triangles and one hexagon, in that order.

\begin{figure}[htbp] % figure placement: here, top, bottom, or page
\centering
\includegraphics[width=3.5in]{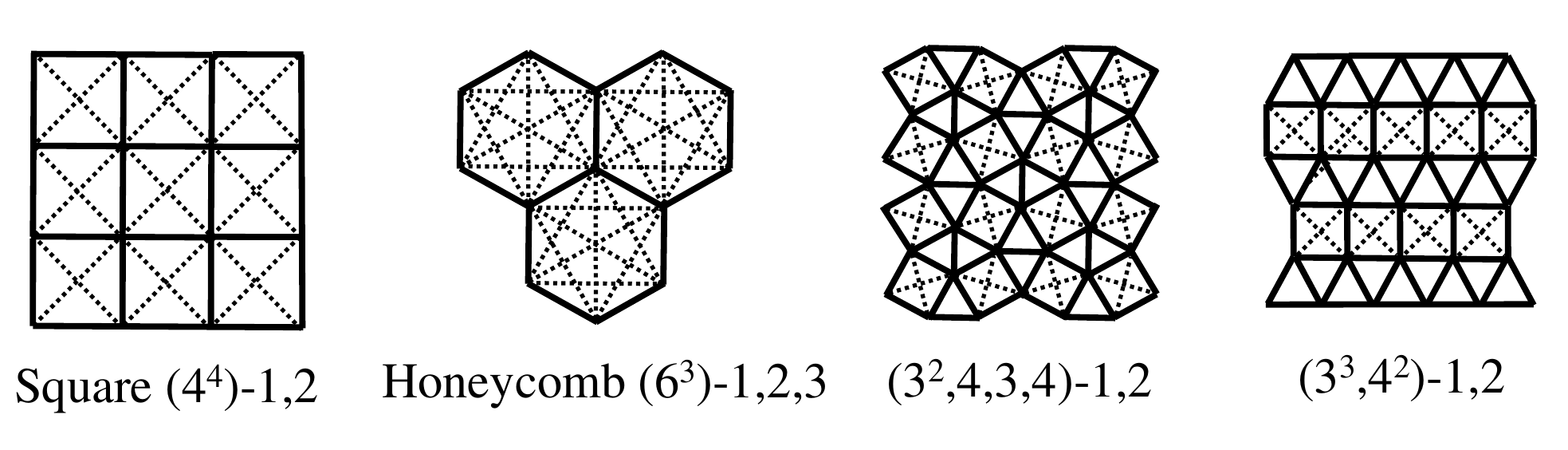}
\caption{Illustration showing that the matching lattices of these four Archimedean lattices are the same lattices with up to the second or third NN. The thresholds of these lattices is one minus the threshold of the original lattices.}
\label{fig:matching}
\end{figure}

In the late 1990's, d'Iribarne, Rasigni and Rasigni \cite{dIribarneRasigniRasigni95,dIribarneRasigniRasigni99,dIribarneRasigniRasigni99b} determined the site percolation thresholds of all Archimedean lattices with extended ranges up to the 10th NN.  We have listed those thresholds in  Table \ref{tab:siteperholds2d}, updated with more precise results in some cases \cite{Malarz21,XunHaoZiff2021}, and precise thresholds for the standard lattices (with first NN).  Furthermore, we can make use of the fact that some of the extended-range lattices are matching lattices of the same lattice with first NN, and therefore have the complementary threshold $1-p_c$, as shown in Fig.\ \ref{fig:matching}. (In a matching lattice, all faces with more than three sides are replaced by a complete graph that connects all pairs of vertices together.)   

\begin{figure}[htbp] %  figure placement: here, top, bottom, or page
\centering
\includegraphics[width=1 in]{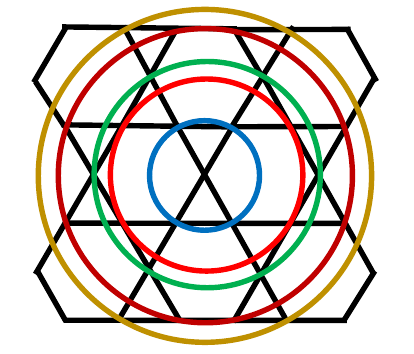}
\caption{Illustration showing the first five shells of nearest neighbors (NN) of the kagome (3,6,3,6) lattice. In the first shell (radius $r=1$), there are 4 NN, in the second shell ($r=\sqrt{3}$), there are 4 NN, in the third shell ($r= 2$), there are 6 NN, in the fourth shell ($r = \sqrt{7}$), there are 8 NN, and in the fifth shell ($r = 3$), there are 4 NN, as listed in Table \ref{tab:siteperholds2d}. Looking at larger $r$, we find that the shell index number appears to grow as a power-law $\sim r^{2.27}$.}
\label{fig:kagomeshells}
\end{figure}

\begin{figure}[htbp] % figure placement: here, top, bottom, or page
\centering
\includegraphics[width=2.5in]{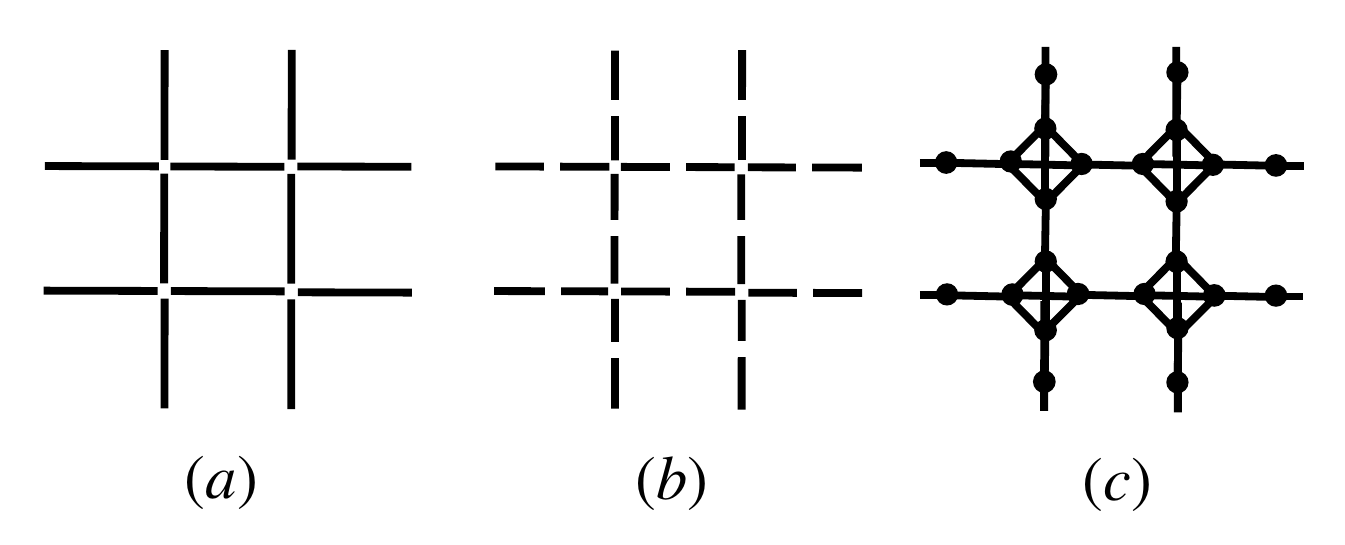}
\caption{Derivation of the threshold of the $(4,8^2)$ lattice with first- and second-NN bonds by the bond-to-site (covering) transformation: (a) Bond percolation on a square lattice, $p_c$(bond) = 1/2.  (b) Bond percolation on the square lattice with double bonds (the Lieb lattice), $p_c$(bond)$ = 1/\sqrt{2}$. (c) Site percolation on the $(4,8^2)$-1,2 lattice, $p_c$(site)$=1/\sqrt{2}$.}
\label{fig:foureight}
\end{figure}

\begin{figure}[htbp] % figure placement: here, top, bottom, or page
\centering
\includegraphics[width=3.5 in]{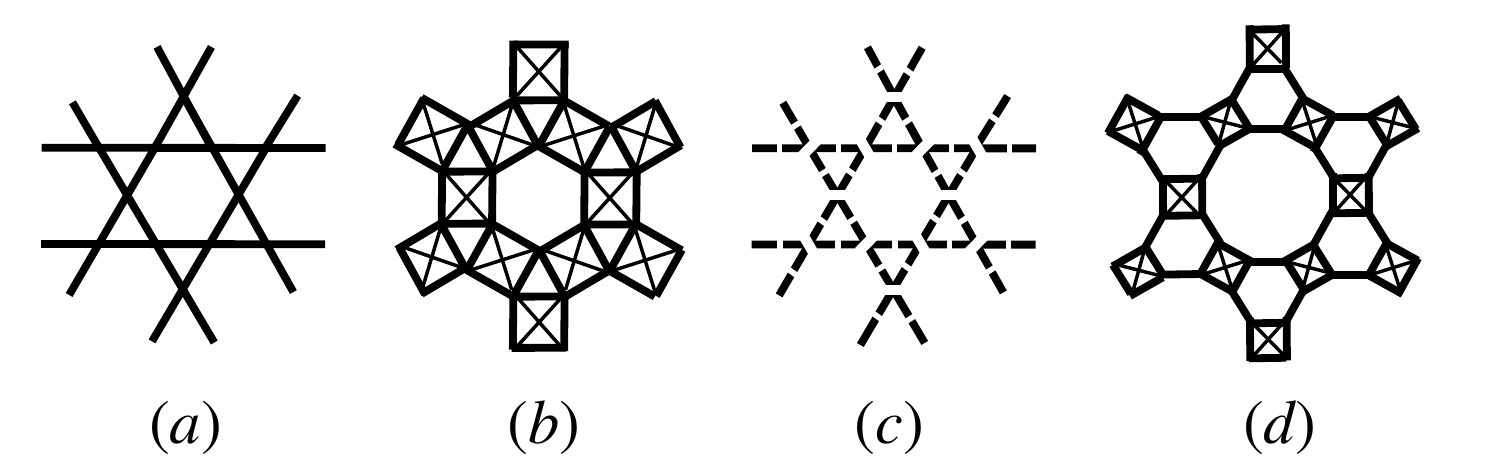}
\caption{(a) Kagome lattice with $p_c$(bond) $=0.524405$ \cite{Jacobsen14}. (b) The covering lattice of (a) is the (3,4,6,4) lattice with first and second NN, with $p_c$(site) = 0.524405. (c) Kagome double-bond lattice with $p_c$(bond)$=(0.524405)^{1/2}=0.724128$. (d) Covering lattice of (c) is the (4,6,12) lattice with first and second NN, with $p_c$(site)$=0.724128$. Note you can go directly from (d) to (b) by replacing two sites along the inner edges of the hexagons by one site, so that the threshold $p_c$ becomes squared, similar to the case of the (3,12$^2$) versus the (6$^3$) lattice \cite{SudingZiff99}. }
\label{fig:kagome}
\end{figure}

We can also find additional results by using the fact that a bond problem can be converted to a site problem by replacing the lattice by the line graph or covering lattice, which connects the centers of the bonds together to create a new lattice.  The $(4,8^2)$ lattice with first and second NN is the covering lattice of the square lattice with double bonds (the Lieb lattice), with each bond having a threshold of $p_c = \sqrt{1/2}$, and consequently this is the site threshold of the $(4,8^2)$ lattice with first and second NN bonds as shown in Fig.\ \ref{fig:foureight}. The covering lattice of the kagome (3,6,3,6) lattice is the (3,4,6,4) lattice with first and second NN, and therefore the bond threshold of the kagome lattice 0.524405 is the site threshold of the (3,4,6,4)-1,2 (Fig.\ \ref{fig:kagome}b).   A similar construction on the kagome lattice with double bonds shows that the site threshold of the (4,6,12) lattice with first and second NN  is equal to the square root of the bond threshold of the kagome lattice, $(0.524405)^{1/2} = 0.724158$ (Fig.\ \ref{fig:kagome}d). These results are all included in Table \ref{tab:siteperholds2d}.  Comparing these improved values to those found by d'Iribarne et al., we find that the latter are accurate to at least two digits, with some variation in the third digit; for example for the (4,6,12)-1,2 lattice, they give 0.720 compared to the value 0.724158 that we find above.   Still, the results of d'Iribarne et al.\ are sufficiently accurate for our discussion of the general behavior of the thresholds.  In Table \ref{tab:siteperholds2d}, we also list NN, the number of nearest neighbors in each shell, the total $z$ up to that shell $z_\mathrm{total}$, and the square of the radius of that shell $r^2$.  An example of the first five shells of neighbors of the kagome (3,6,3,6) lattice is shown in Fig.\ \ref{fig:kagomeshells}.

\begin{figure}[htbp] % figure placement: here, top, bottom, or page
\centering
\includegraphics[width=3.2in]{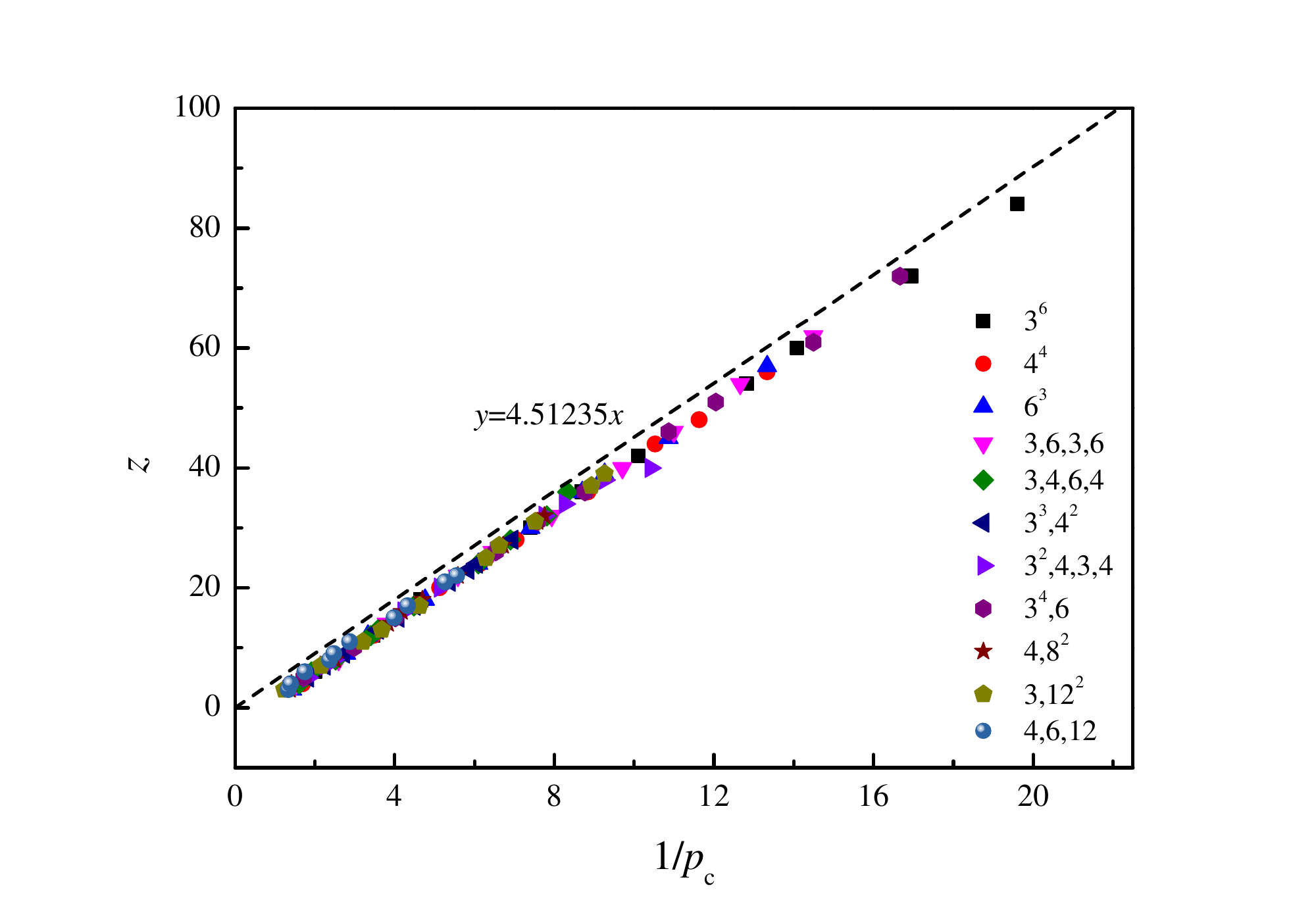}
\caption{Plots of $z$ vs $1/p_{c}$ for the 11 Archimedean lattices with various ranges of NN, using the data of Table \ref{tab:siteperholds2d}.   The virtual straight line is a guideline with a slope of $4.512$ and intercept of $0$, corresponding to Eq.\ (\ref{eq:approx2d}).}
\label{fig:z-vs-pc-site-2d}
\end{figure}

\begin{figure}[htbp] % figure placement: here, top, bottom, or page
\centering
\includegraphics[width=3.2in]{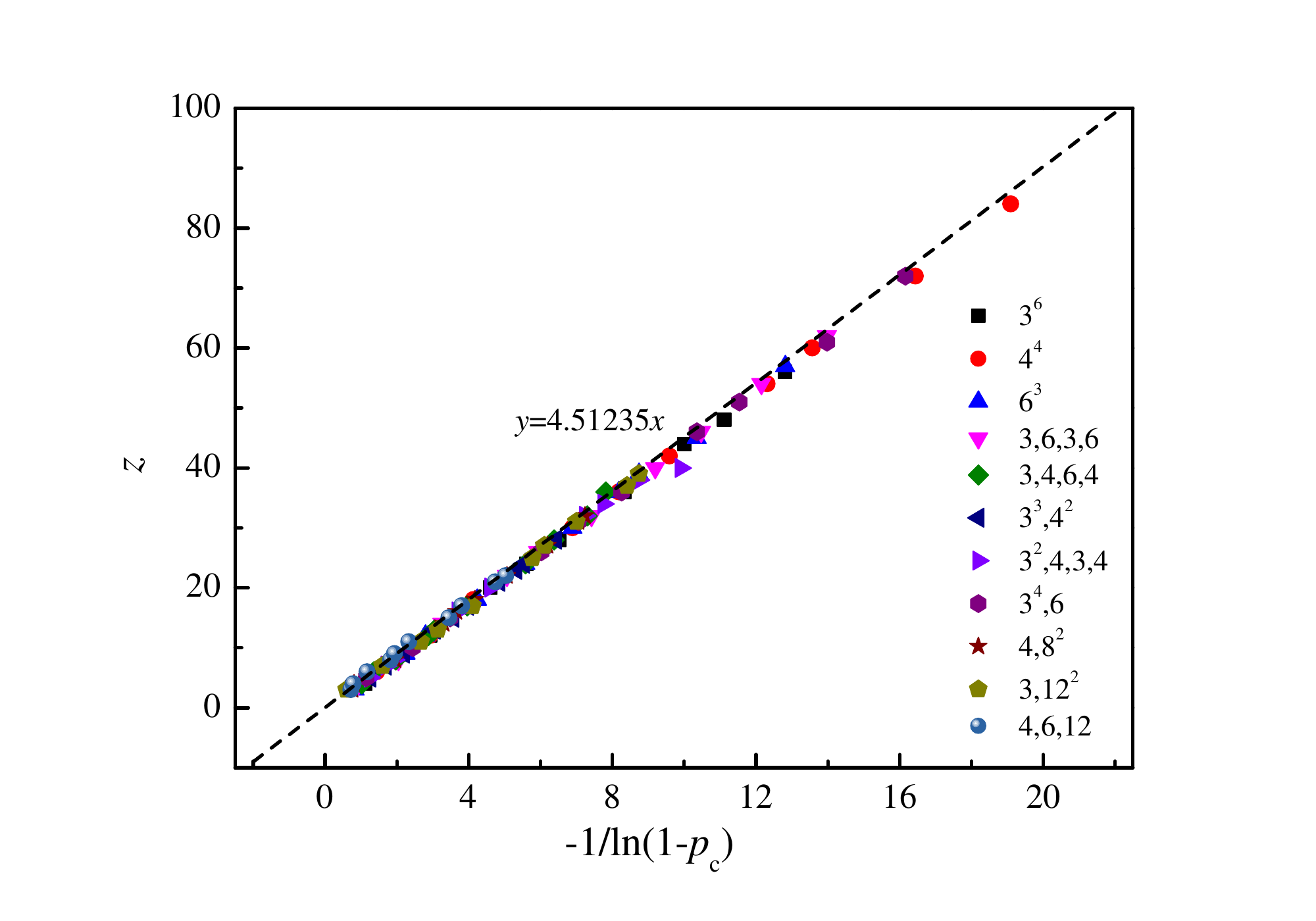}
\caption{Plot of $z$ versus $-1/\ln(1-p_{c})$ for the 11 Archimedean lattices with various ranges of NN. The virtual straight line is a guideline with a slope of $4.512$ and intercept of $0$, corresponding to Eq.\ (\ref{eq:approx2d}).}
\label{fig:z-vs-ln(1-pc)-site-2d}
\end{figure}

Using the data of Table \ref{tab:siteperholds2d}, we plot the behavior of $z$ versus $1/p_{c}$ and $z$ versus $-1/\ln(1-p_{c})$, proposed by Eqs.\ (\ref{eq:z}) and (\ref{eq:lnpc}), for the different lattices, in Figs.\ \ref{fig:z-vs-pc-site-2d} and \ref{fig:z-vs-ln(1-pc)-site-2d}, respectively. In these plots, the discrete points represent the data for each lattice, and the virtual straight line is a guideline with a slope of $4.512$ and intercept of zero representing Eq.\ (\ref{eq:approx2d}). It is seen that nearly all the lattices collapse to the same line, and most of the plots give a slope near the predicted value of $4.512$. This result demonstrates the universality of asymptotic behavior for $zp_c$ for site percolation on all extended-range Archimedean lattices over a wide range of $z$.

While both plots have a slope near the predicted value from $\eta_c$, the points of Fig.\ \ref{fig:z-vs-ln(1-pc)-site-2d} fit well the assumption of an intercept equal to zero, while those of Fig.\ \ref{fig:z-vs-pc-site-2d} fit to a line with the correct slope but a non-zero intercept, which corresponds to the constant $-b$ in Eq.\ (\ref{eq:z}).  The data for the various lattices predicts $b$ in the range  $2.134$ to $3.435$, with an average value of $\approx 3.0$.  By using Eq.\ (\ref{eq:lnpc}) rather than Eq.\ (\ref{eq:z}) to fit the thresholds, we get a satisfactory fit of the finite-$z$ behavior without the need of an additional parameter.  

We can also look at different lattices that share the same value of $z = z_\mathrm{total}$ and compare their thresholds.  For example, $z=12$ corresponds to the {\sc tri}-1,2 lattice ($p_c = 0.29026$), the {\sc sq}-3 lattice ($p_c = 0.289123$), the {\sc hc}-1,2,3 lattice ($p_c = 0.302960$), and the (3,4,6,4)-1,2,3,4 lattice ($p_c = 0.294$), where the numbers after the lattice name represents the NN.  Notice that the thresholds are close together.  Our theoretical formulas give $p_c = 0.37602$ (Eq.\ (\ref{eq:approx2d})), $p_c = 0.30082$ (Eq.\ (\ref{eq:z}) with $b = 3$), and $p_c = 0.31341$ (Eq.\ (\ref{eq:general})).  Clearly, Eq.\ (\ref{eq:z}) with $b=3$ gives the best approximation to the actual thresholds.

For a larger $z = 36$, we find the thresholds also close together:  {\sc tri}-1,2,3,4,5 ($p_c = 0.11574$), {\sc sq}-1,...,7 ($p_c = 0.115348$),  {\sc hc}-1,...,7 ($p_c = 0.115$), (3,4,6,4)-1,...,10 ($p_c = 0.120$), and (3$^4$,6)-1,...,6 ($p_c = 0.114$).  The approximation formulas give $p_c = 0.12534$ (Eq.\ (\ref{eq:approx2d})), $p_c = 0.11570$ (Eq.\ (\ref{eq:z}) with $b = 3$), and $p_c = 0.11780$ (Eq.\ (\ref{eq:general})).  Once again, Eq.\ (\ref{eq:z}) with $b=3$ gives the best approximation.  To make a better comparison with theory, it would be useful to have the results of the last two lattices above to higher precision.

% In addition, the asymptotic behavior is essentially independent of the shape of the neighborhood, that is, does not depend upon the type of lattice. The coordination number $z$ appears to be the principal determinant of the thresholds of systems with extended neighborhoods, with the lattice type and neighborhood shape seemingly less important. 

%One can roughly judge the reliability of percolation thresholds based on these properties. In Fig. \ref{fig:z-vs-pc-site-2d}, there is a point, that is the 10th nearest neighbor for $3^{2}4.3.4$ lattice and indicated by the arrow, deviating from the straight line. We conjecture this site percolation threshold should be $0.10...$ instead of $0.096$. ({\color{red} Numerical simulation is needed to prove this conjecture?}) 

% Fig. \ref{fig:z-vs-pc-site-2d} and Fig. \ref{fig:z-vs-ln(1-pc)-site-2d}, which one is better? Further comparison is carried out, and we find: (1) Fig. \ref{fig:z-vs-pc-site-2d} leads to the slope more closer to the predicted value of $4.51235$ than Fig. \ref{fig:z-vs-ln(1-pc)-site-2d}, (2) while the intercept is closer to $0$ when we plot $z$ versus $-1/ln(1-p_{c})$. We will make a quantitative analysis about these differences in the next section about site percolation in 3D.

\begin{table*}[htbp]
%\scriptsize

\caption{Site percolation thresholds and other properties, including $r^{2}$, NN, and total $z$, for the 11 Archimedean lattices with extended neighborhoods, up to the 10th NN. Values of $p_{c}$ shown to three significant digits are from Ref.\ \cite{dIribarneRasigniRasigni99b}, while higher precision results (many truncated) are from $^\mathrm{a}$Ref.\ \cite{Malarz21}, $^\mathrm{b}$Ref.\ \cite{Jacobsen14}, $^\mathrm{c}$Ref.\ \cite{XunHaoZiff2021}, $^\mathrm{d}$Ref.\ \cite{SudingZiff99}, $^\mathrm{e}$exact, $^\mathrm{f}$by matching, $^\mathrm{g}$by bond-to-site transformation.} %$^\mathrm{d}$Ref.\ \cite{FengDengBlote08} 
\begin{tabular}{cccccccccccc}
\hline\hline
    Lattice    &             &    \multicolumn{10}{c}{Nearest-neighbor number}   \\
\cline{3-12}
                &             & 1     & 2     & 3            & 4     & 5            & 6     & 7     & 8     & 9     & 10      \\  \hline 
    $(3^{6})$   & $r^{2}$     & 1     & 3     & 4            & 7     & 9            & 12    & 13    & 16    & 19    & 21      \\
    {\sc tri}         & NN          & 6     & 6     & 6            & 12    & 6            & 6     & 12    & 6     & 12    & 12      \\
                & total $z$   & 6     & 12    & 18           & 30    & 36           & 42    & 54    & 60    & 72    & 84      \\
                & $p_{c}$     & 0.500000$^\mathrm{e}$ & 0.29026$^\mathrm{a}$ & 0.21546$^\mathrm{a}$ & 0.13582$^\mathrm{a}$ & 0.11574$^\mathrm{a}$ & 0.099 & 0.078 & 0.071 & 0.059 & 0.051   \\ 
                &             &       &               &               &              &              &              &              &              &       &         \\   
    $(4^{4})$   & $r^{2}$     & 1     & 2             & 4             & 5            & 8            & 9            & 10           & 13           & 16    & 17      \\
    {\sc sq}          & NN          & 4     & 4             & 4             & 8            & 4            & 4            & 8            & 8            & 4     & 8       \\
                & total $z$   & 4     & 8             & 12            & 20           & 24           & 28           & 36           & 44           & 48    & 56      \\
                & $p_{c}$     & 0.592746$^\mathrm{b}$ & 0.407254$^\mathrm{f}$  & 289123$0.^\mathrm{c}$ & 0.196729$^\mathrm{c}$ & 0.164712$^\mathrm{c}$ & 0.143255$^\mathrm{c}$ & 0.115348$^\mathrm{c}$ & 0.095766$^\mathrm{c}$ & 0.086 & 0.075   \\
                &             &       &       &       &       &       &       &       &       &       &         \\             
    $(6^{3})$   & $r^{2}$     & 1     & 3     & 4     & 7     & 9     & 12    & 13    & 16    & 19    & 21      \\
     {\sc hc}            & NN          & 3     & 6     & 3     & 6     & 6     & 6     & 6     & 3     & 6     & 12      \\
                & total $z$   & 3     & 9     & 12    & 18    & 24    & 30    & 36    & 39    & 45    & 57      \\
                & $p_{c}$     & 0.697040$^\mathrm{b}$ & 0.359 & 0.302960$^\mathrm{f}$ & 0.210 & 0.164 & 0.135 & 0.115 & 0.108 & 0.092 & 0.075   \\
                &             &       &       &       &       &       &       &       &       &       &         \\              
    $(3,6,3,6)$ & $r^{2}$     & 1     & 3     & 4     & 7     & 9     & 12    & 13    & 16    & 19    &  21     \\
      {\sc kag}           & NN          & 4     & 4     & 6     & 8     & 4     & 6     & 8     & 6     & 8     & 8       \\
                & total $z$   & 4     & 8     & 14    & 22    & 26    & 32    & 40    & 46    & 54    & 62      \\
                & $p_{c}$     & $0.652703^\mathrm{e}$ & 0.386 & 0.263 & 0.179 & 0.155 & 0.126 & 0.103 & 0.091 & 0.079 & 0.069   \\
                &             &       &       &       &       &       &       &       &       &       &         \\
    $(3,4,6,4)$ & $r^{2}$     & 1     & 2     & 3     & 2+$\sqrt{3}$  & 4     & 4+$\sqrt{3}$ & 4+2$\sqrt{3}$ & 5+2$\sqrt{3}$ & 6+3$\sqrt{3}$ & 8+2$\sqrt{3}$ \\
                & NN          & 4     & 2     & 2     & 4             & 1     & 4            & 7             & 4             & 4             & 4             \\
                & total $z$   & 4     & 6     & 8     & 12            & 13    & 17           & 24            & 28            & 32            & 36            \\
                & $p_{c}$     & $0.621812^\mathrm{b}$ & $0.524405^\mathrm{b,g}$ & 0.398 & 0.294         & 0.279 & 0.223        & 0.164         & 0.145         & 0.128         & 0.120         \\
                &             &       &       &       &               &       &              &               &               &               &               \\
  $(3^{3},4^{2})$ & $r^{2}$     & 1     & 2     & 3     & 2+$\sqrt{3}$ & 4     & 5     & 4+$\sqrt{3}$ & 7     & 4+2$\sqrt{3}$ & 5+2$\sqrt{3}$  \\
                & NN          & 5     & 2     & 2     & 4            & 2     & 2     & 4            & 2     & 1             & 4              \\
                & total $z$   & 5     & 7     & 9     & 13           & 15    & 17    & 21           & 23    & 24            & 28             \\
                & $p_{c}$     & $0.550213^\mathrm{d}$ & $0.449787^\mathrm{f}$ & 0.366 & 0.279        & 0.244 & 0.222 & 0.186        & 0.171 & 0.165         & 0.144          \\
                &             &       &       &       &              &       &       &              &       &               &                \\
  $(3^{2},4,3,4)$ & $r^{2}$     & 1     & 2     & 3     & 2+$\sqrt{3}$ & 4+$\sqrt{3}$ & 4+2$\sqrt{3}$ & 5+2$\sqrt{3}$ & 7+2$\sqrt{3}$ & 6+3$\sqrt{3}$ & 8+2$\sqrt{3}$ \\
                & NN          & 5     & 2     & 1     & 8            & 4            & 6             & 6             & 2             & 4             & 2             \\
                & total $z$   & 5     & 7     & 8     & 16           & 20           & 26            & 32            & 34            & 38            & 40            \\
                & $p_{c}$     & $0.550806^\mathrm{d}$ & $0.449104^\mathrm{f}$ & 0.405 & 0.237        & 0.195       & 0.153         & 0.129         & 0.121         & 0.108         & 0.096         \\
                &             &       &       &       &              &              &               &               &               &               &               \\
    $(3^{4},6)$   & $r^{2}$     & 1     & 3     & 4     & 7     & 9     & 12    & 13    & 16    & 19    & 21     \\
                & NN          & 5     & 5     & 5     & 11    & 5     & 5     & 10    & 5     & 10    & 11     \\
                & total $z$   & 5     & 10    & 15    & 26    & 31    & 36    & 46    & 51    & 61    & 72     \\
                & $p_{c}$     &  $0.579498^\mathrm{b}$ & 0.335 & 0.249 & 0.153 & 0.132 & 0.114 & 0.092 & 0.083 & 0.069 & 0.060  \\
                &             &       &       &       &       &       &       &       &       &       &        \\
    $(4,8^{2})$   & $r^{2}$     & 1     & 2     &2+$\sqrt{2}$&3+2$\sqrt{2}$&4+2$\sqrt{2}$&5+2$\sqrt{2}$&6+3$\sqrt{2}$&6+4$\sqrt{2}$&7+4$\sqrt{2}$&9+4$\sqrt{2}$  \\
                & NN          & 3     & 1     & 4          & 6           & 2           & 2           & 4           & 5           & 4           & 1             \\
                & total $z$   & 3     & 4     & 8          & 14          & 16          & 18          & 22          & 27          & 31          & 32            \\
                & $p_{c}$     & $0.729723^\mathrm{b}$ & $0.707107^\mathrm{e,g}$ & 0.399      & 0.261       & 0.239       & 0.213       & 0.179       & 0.149       & 0.132       & 0.129         \\
                &             &       &       &            &             &             &             &             &             &             &               \\
        $(3,12^{2})$  & $r^{2}$     & 1     &2+$\sqrt{3}$&4+2$\sqrt{3}$&5+2$\sqrt{3}$&6+3$\sqrt{3}$&7+4$\sqrt{3}$&8+4$\sqrt{3}$&10+5$\sqrt{3}$&11+6$\sqrt{3}$&12+6$\sqrt{3}$\\
                & NN          & 3     & 4          & 4           & 2           & 4           & 8           & 2           & 4            & 6            & 2            \\
                & total $z$   & 3     & 7          & 11          & 13          & 17          & 25          & 27          & 31           & 37           & 39           \\
                & $p_{c}$     & 0.807901$^\mathrm{e}$ & 0.464      & 0.312       & 0.272       & 0.216       & 0.159       & 0.151       & 0.133        & 0.112        & 0.108      \\        
                          &             &       &       &            &             &             &             &             &             &             &   \\
             $(4,6,12)$    & $r^{2}$     & 1     & 2     & 3     & 2+$\sqrt{3}$ & 4     & 4+$\sqrt{3}$ & 4+2$\sqrt{3}$ & 5+2$\sqrt{3}$ & 6+3$\sqrt{3}$ & 8+2$\sqrt{3}$    \\
                & NN          & 3     & 1     & 2     & 2            & 1     & 2            & 4             & 2             & 4             & 1                \\
                & total $z$   & 3     & 4     & 6     & 8            & 9     & 11           & 15            & 17            & 21            & 22               \\
                & $p_{c}$     &  0.747801$^\mathrm{b,g}$ & 0.724158$^\mathrm{g}$ & 0.571 & 0.421        & 0.403 & 0.348        & 0.251         & 0.231         & 0.190         & 0.180  \\
\hline\hline
\end{tabular}
\label{tab:siteperholds2d}
\end{table*}

\section{Site and bond percolation on sc, bcc, and fcc lattices with extended connections}
\label{sec:3d}
Here we carried out extensive Monte-Carlo simulations, using a single-cluster growth method \cite{LorenzZiff1998,XunZiff2020,XunZiff2020b}. In this method, many individual clusters are generated from a seeded site on the lattice, and clusters with different sizes $s$ are put in different bins with the range of $(2^n, 2^{n+1}-1)$ for $n=0,1,2,\ldots$. Clusters still growing when they reach an upper size cutoff (this value is set to avoid wrapping around the boundaries and to limit the run time) are counted in the last bin.  Define $n_s(p)$ as the number of clusters (per site) containing $s$ occupied sites as a function of the site or bond occupation probability $p$.  In the scaling limit, in which $s$ is large and $(p-p_{c})$ is small such that $(p-p_{c}) s^ \sigma$ is constant, $n_{s}(p)$ behaves as
\begin{equation}
    n_{s}(p) \sim A_0 s^{-\tau} f[B_0 (p-p_{c}) s^ \sigma],
    \label{eq:nsp}
\end{equation}
where $\tau$, $\sigma$, and $f(x)$ are universal (having same values in a given dimension), while $A_0$ and $B_0$ are lattice-dependent factors. The probability that a point belongs to a cluster of size greater than or equal to $s$ is given by $P_{\ge s} = \sum_{s'=s}^\infty s' n_{s'}$, and it follows for large $s$ and small $(p-p_{c}) s^ \sigma$ that $ s^{\tau - 2} P_{\geq s}$ behaves as
\begin{equation}
    s^{\tau - 2} P_{\geq s} \sim A_1 + B_1 (p-p_{c}) s^ \sigma+C_1 s^{-\Omega},
    \label{eq:nsp2}
\end{equation}
Here we also added a correction-to-scaling term with exponent $\Omega$.  The $A_1$, $B_1$ and $C_1$ are non-universal constants. From the behavior of this quantity, we can easily determine if we are above, near, or below the percolation threshold. One can see Refs.\ \cite{LorenzZiff1998,XunZiff2020,XunZiff2020b} for more details about the single-cluster method.

With regard to the universal exponents $\tau$, $\Omega$, and $\sigma$, in three dimensions, relatively accurate and acceptable results are known: $2.18906(8)$ \cite{BallesterosFernandezMartin-MayorSudupeParisiRuiz-Lorenzo1997}, $2.18909(5)$ \cite{XuWangLvDeng2014} for $\tau$, $0.64(2)$ \cite{LorenzZiff1998}, $0.65(2)$ \cite{GimelNicolaiDurand2000}, $0.60(8)$ \cite{Tiggemann2001}, $0.64(5)$ \cite{BallesterosFernandezMartin-MayorSudupeParisiRuiz-Lorenzo1999} for $\Omega$, and $0.4522(8)$ \cite{BallesterosFernandezMartin-MayorSudupeParisiRuiz-Lorenzo1997}, $0.45237(8)$ \cite{XuWangLvDeng2014}, $0.4419$ \cite{Gracey2015} for $\sigma$. In our simulations, $\tau = 2.18905(15)$, $\Omega = 0.63(4)$, and $\sigma = 0.4522(2)$ are chosen. Here we take large error bars on these values for the sake of safety.

The upper size cutoff is set to be $s_\mathrm{max} = 2^{16}$ occupied sites. Monte-Carlo simulations are performed on systems of size $L\times L \times L$ with $L=512$ under periodic boundary conditions (although the boundaries are actually never reached). Some $10^9$ independent samples were produced for \textsc{bcc} and \textsc{fcc} lattices with 2nd and 3rd NN, and $10^8$ for the \textsc{sc} lattice with $n$th NN, with $n$ from 5 to 13. The number of clusters greater than or equal to size $s$ could be found based on the data from our simulations, and the quantity $s^{\tau-2}P_{\geq s}$ could easily be calculated.

\subsection{Site percolation}
\label{sec:site3d}
We use the notation $\textsc{bcc}$-$a,b,\ldots$ to indicate a \textsc{bcc} lattice with the $a$-th NN, the $b$-th NN, etc., and similarly for the \textsc{fcc} and \textsc{sc} lattices. 

\begin{figure}[htbp] %  figure placement: here, top, bottom, or page
\centering
\includegraphics[width=3.3in]{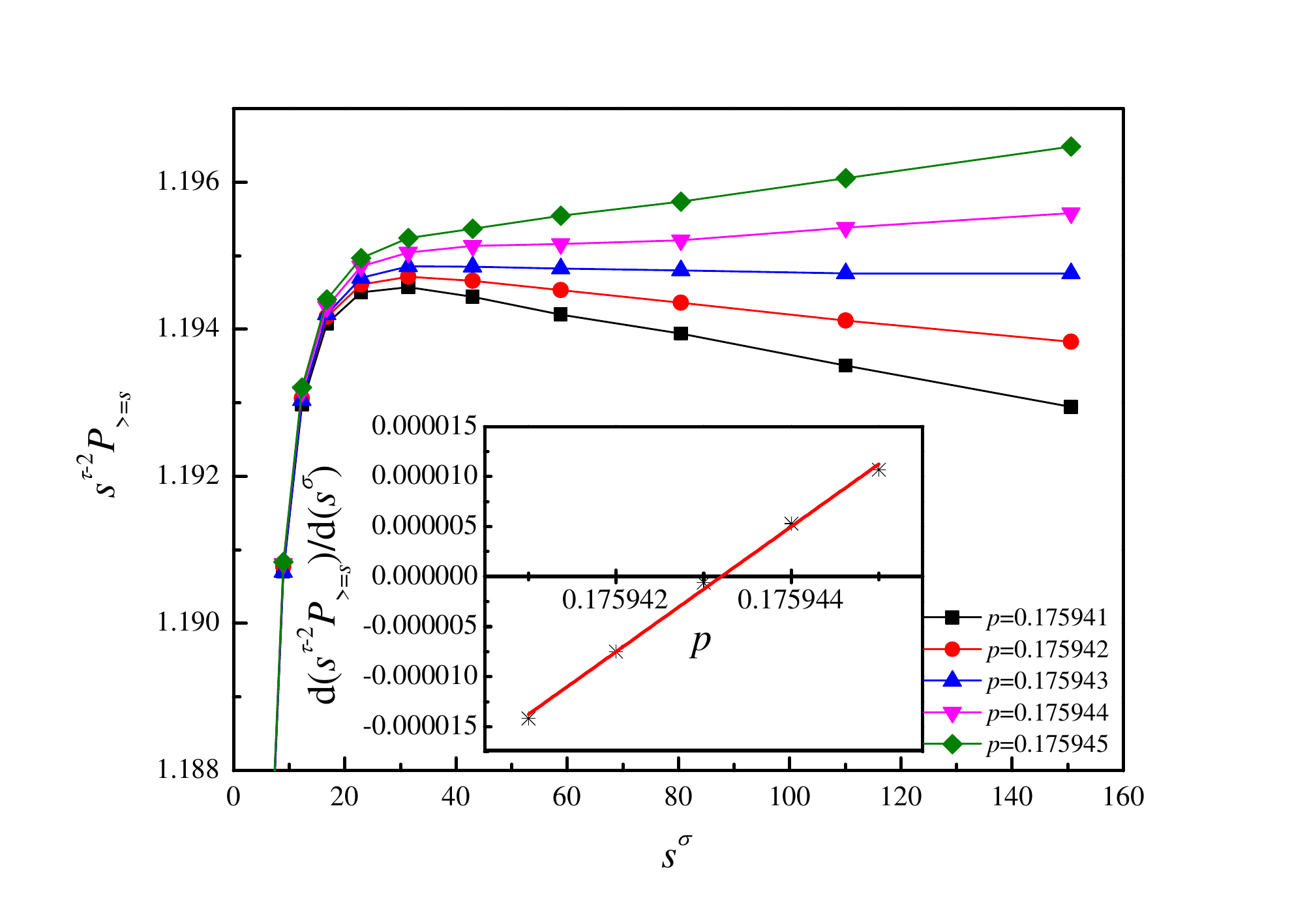}
\caption{Plot of $s^{\tau-2}P_{\geq s}$ versus $s^{\sigma}$ with $\tau = 2.18905$ and $\sigma = 0.4522$ for site percolation of the \textsc{bcc}-1,2 lattice under different values of $p$. The inset indicates the slope of the linear portions of the curves shown in the main figure as a function of $p$, and the predicted value of $p_{c} = 0.1759432$ can be calculated from the $p$ intercept.}
\label{fig:bcc-1-2-sigma-site}
\end{figure}

Figure \ref{fig:bcc-1-2-sigma-site} shows the relation of $s^{\tau-2}P_{\geq s}$ versus\ $s^{\sigma}$ %with $\tau = 2.18905$ and $\sigma = 0.4522$
for site percolation of the \textsc{bcc}-1,2 lattice, under probabilities $p = 0.175941$, $0.175942$, $0.175943$, $0.175944$ and $0.175945$. (Preliminary work narrowed the search to this range.) For small clusters, we can see a steep rise of  $s^{\tau-2}P_{\geq s}$ corresponding to the finite-size-effect term, $s^{-\Omega}$, while for large clusters, the plot shows linear behavior as expected from Eq.\ (\ref{eq:nsp2}). The linear portion of the curve become more nearly horizontal when $p$ is close to $p_c$. The estimated value of $p_c$ can then be deduced using $\mathrm{d} (s^{\tau-2}P_{\geq s})/ \mathrm{d} (s^{\sigma}) \sim B_1(p-p_{c})$, as shown in the inset of Fig.\ \ref{fig:bcc-1-2-sigma-site}, where $p_{c} = 0.1759432$ can be estimated from the $p$ intercept of the plot of the derivative versus $p$.

\begin{figure}[htbp] %  figure placement: here, top, bottom, or page
\centering
\includegraphics[width=3.3in]{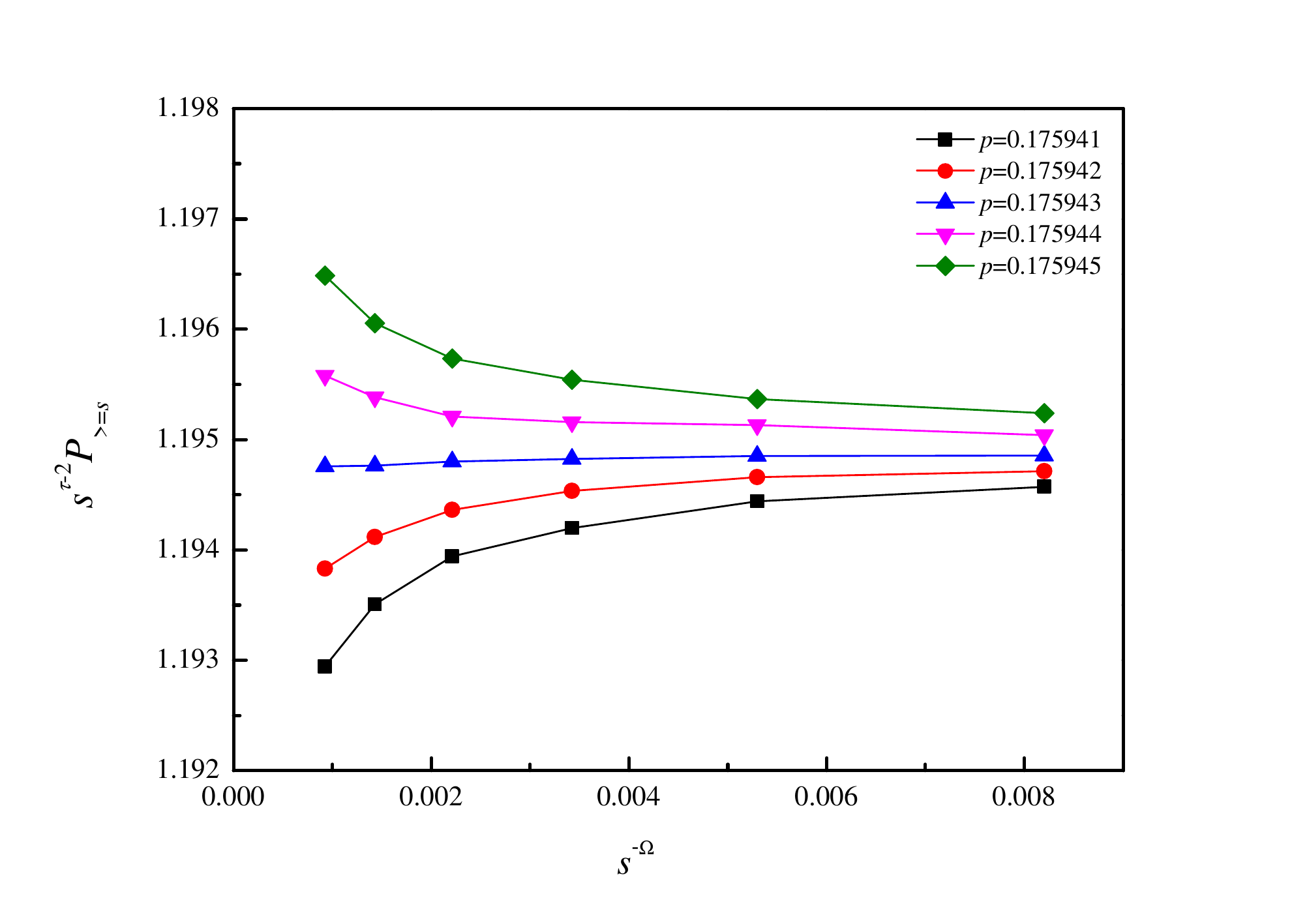}
\caption{Plot of $s^{\tau-2}P_{\geq s}$ versus $s^{-\Omega}$ with $\tau = 2.18905$ and $\Omega = 0.63$ for site percolation of the \textsc{bcc}-1,2 lattice under different values of $p$.}
\label{fig:bcc-1-2-omega-site}
\end{figure}

When $p$ is close to $p_c$, a plot of $s^{\tau-2}P_{\geq s}$ versus $s^{-\Omega}$ is useful to estimate the percolation threshold. Figure \ref{fig:bcc-1-2-omega-site} shows this plot for the \textsc{bcc}-1,2 lattice under probabilities $p = 0.175941$, $0.175942$, $0.175943$, $0.175944$, and $0.175945$. Linear behavior for large $s$ (small abscissa) can be seen when $p$ is very close to $p_{c}$, while when $p$ is away from $p_c$, the curves show obvious deviations from linearity for large $s$. Based on these curves, the range $0.175943 < p_{c} < 0.175944$ can be concluded, which is consistent with the value we deduced from Fig.\ \ref{fig:bcc-1-2-sigma-site}, $p_{c} = 0.1759432$.  

Thus, we conclude that the site percolation threshold of the \textsc{bcc}-1,2 lattice to be $p_{c} = 0.1759432(8)$, where the number in parentheses represents the estimated error in the last digit, by comprehensively considering the two methods mentioned above, as well as the errors for the values of $\tau$, $\Omega$ and $\sigma$. The simulation results for the other three lattices we considered  (\textsc{bcc}-1,2,3, \textsc{fcc}-1,2, and \textsc{fcc}-1,2,3) are shown in the Supplementary Material \cite{XunZiff2021supplementary} in Figs.\ 1-6, and the corresponding thresholds are summarized in Table \ref{tab:siteperholds3d}. Previously reported results are also shown, and it can be seen that the accuracy of the thresholds has been greatly increased.  Our results are generally consistent with previous works, except for the case of the \textsc{sc}-1,2,3,4 lattice where a previous result from Ref.\ \cite{Malarz2015} was evidently in error \cite{XunZiff2020}.

%\setcitestyle{super,open={},close={}}
\begin{table}[htb]
\caption{Site percolation thresholds for the \textsc{sc}, \textsc{bcc}, and \textsc{fcc} lattices with various ranges of NN.  The \textsc{bcc} and \textsc{fcc} results were determined here, while the  results for the  \textsc{sc} lattice come from Ref.\ \cite{XunHaoZiff2021}. Previous results are also shown.}
\begin{tabular}{ccccl}
\hline\hline
    lattice                     & $z$   & $p_{c}$       &  $zp_{c}$& previous values \\ \hline
    \textsc{bcc}-1,2            & 14    & 0.1759432(8) & 2.4632  & 0.175\cite{DaltonDombSykes64}, 0.1686(20)\cite{JerauldScrivenDavis1984}  \\
    \textsc{sc}-1,2             & 18    & 0.1373045(5)  &  2.4715 &  0.137\cite{DaltonDombSykes64}, 0.136\cite{GawronCieplak91}  \\
    & & & & 0.1372(1)\cite{KurzawskiMalarz2012} \\
    \textsc{fcc}-1,2            & 18    & 0.1361408(8)  & 2.4505 & 0.136\cite{DaltonDombSykes64} \\
    \textsc{sc}-1,2,3           & 26    & 0.0976444(6)  & 2.5388 & 0.097\cite{DaltonDombSykes64}, 0.0976(1)\cite{KurzawskiMalarz2012} \\
    \textsc{bcc}-1,2,3          & 26    & 0.0959084(6)  & 2.4936 & 0.095\cite{DombDalton1966} \\ 
    \textsc{sc}-1,2,3,4         & 32    & 0.0801171(9)  & 2.5637 &  0.10000(12)\cite{Malarz2015}\\
    \textsc{fcc}-1,2,3          & 42    & 0.0618842(8)  & 2.5991 & 0.061\cite{DombDalton1966}, 0.0610(5)\cite{GawronCieplak91} \\ 
    \textsc{sc}-1,...,5         & 56    & 0.0461815(5)  & 2.5861 &  \\
    \textsc{sc}-1,...,6         & 80    & 0.0337049(9)  & 2.6964 &  \\
    \textsc{sc}-1,...,7         & 92    & 0.0290800(10) & 2.6754 &  \\ 
    \textsc{sc}-1,...,8         & 122   & 0.0218686(6)  & 2.6680 &  \\ 
    \textsc{sc}-1,...,9         & 146   & 0.0184060(10) &  2.6873 &  \\
\hline\hline
\end{tabular}
\label{tab:siteperholds3d}
\end{table}

%Precise site percolation thresholds for the \textsc{bcc} and the \textsc{fcc} lattices with 2nd and 3rd nearest neighbors are obtained.

\begin{figure}[htbp] %  figure placement: here, top, bottom, or page
\centering
\includegraphics[width=3.3in]{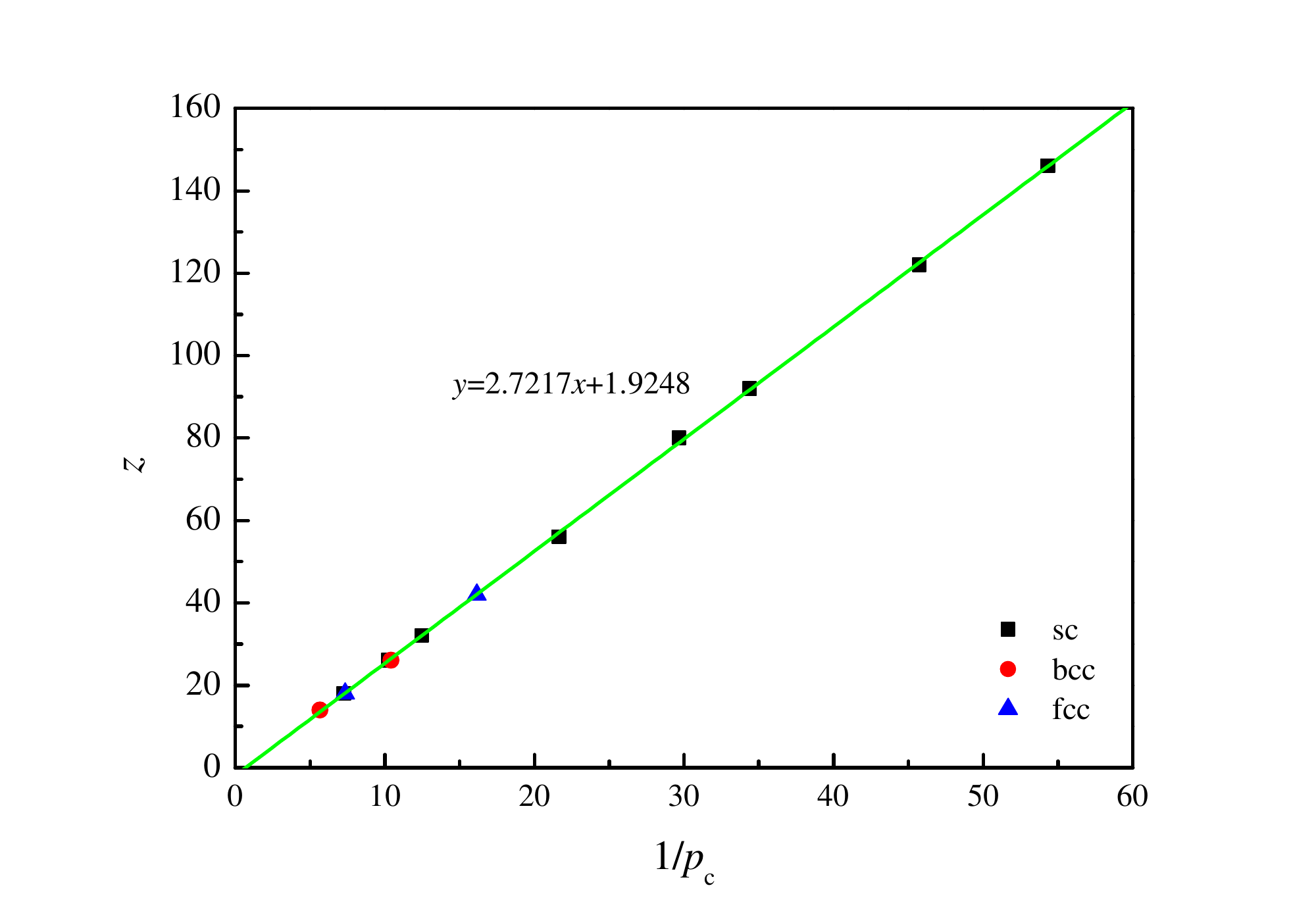}
\caption{Plot of $z$ vs $1/p_{c}$ for the three-dimensional lattices shown in Table \ref{tab:siteperholds3d}. The green line is a fit through the data, and has a slope 2.7217, close to the predicted value 2.7351 from Eq.\ (\ref{eq:approx3d}).}
\label{fig:z-vs-pc-site-3d}
\end{figure}

\begin{figure}[htbp] %  figure placement: here, top, bottom, or page
\centering
\includegraphics[width=3.3in]{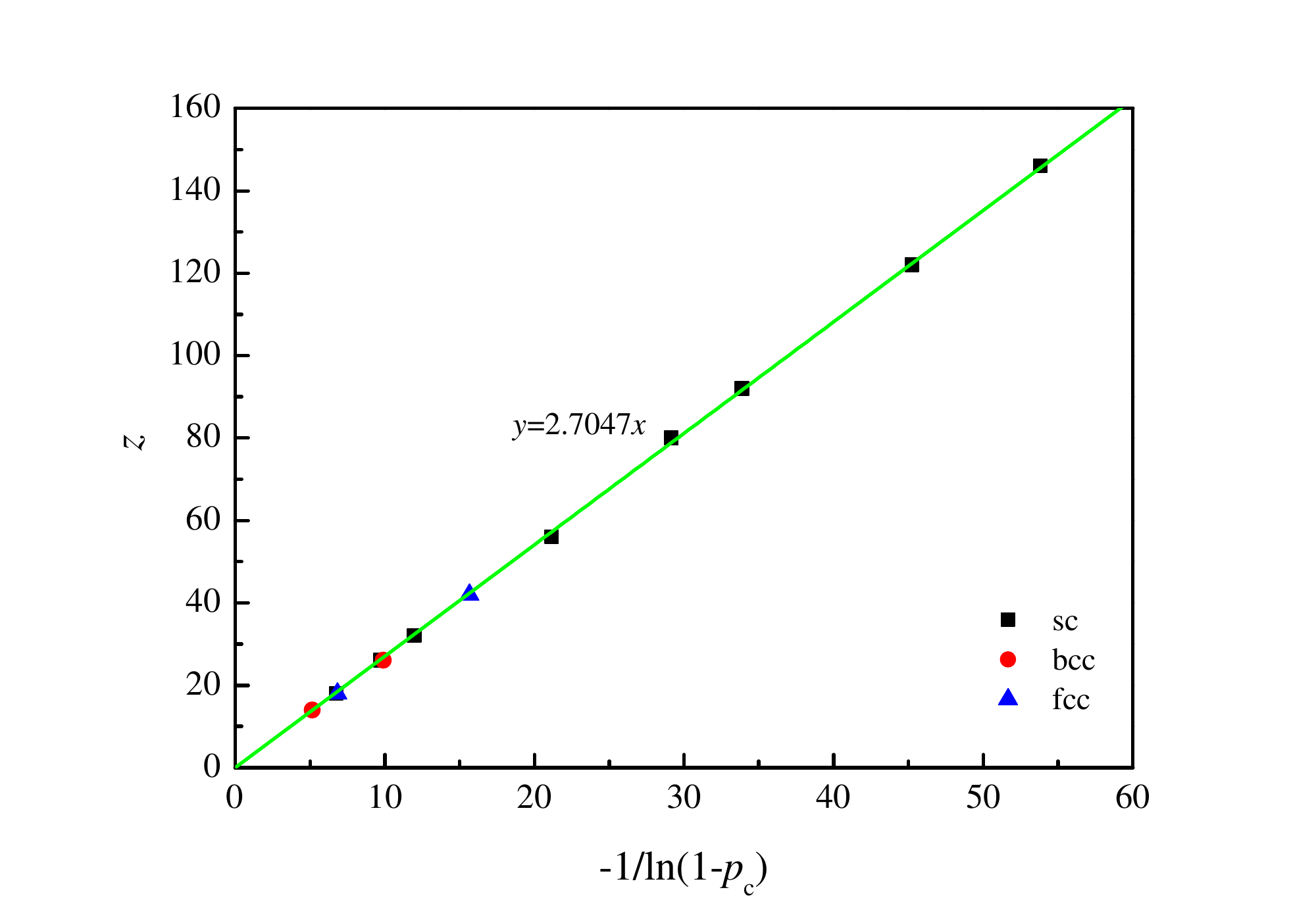}
\caption{Plot of $z$ vs $-1/\ln(1-p_{c})$ for site percolation threshold for the three-dimensional lattices listed in Table \ref{tab:siteperholds3d}.  The green line is a fit through the data, and has a slope 2.7047, close to the predicted value 2.7351 from Eq.\ (\ref{eq:approx3d}).}
\label{fig:z-vs-ln(1-pc)-site-3d}
\end{figure}

For the convenience of the following analysis, in Table \ref{tab:siteperholds3d}, we also show our former site percolation thresholds for \textsc{sc}-1,...,$n$ ($2 \le n \le 9$) lattices  \cite{XunHaoZiff2021}. In addition, the values of $zp_{c}$ are also shown in the fourth column of Table \ref{tab:siteperholds3d}. With the increase of coordination number $z$, the value of $zp_{c}$ shows a gradual increase. Further investigations are performed by plotting the relation of $z$ versus $1/p_{c}$ and $z$ versus $-1/\ln(1-p_{c})$, as shown in Figs.\ \ref{fig:z-vs-pc-site-3d} and \ref{fig:z-vs-ln(1-pc)-site-3d}, respectively. One can see that the results of different lattices collapse onto a line, which has a slope of $2.7217$ for Fig.\ \ref{fig:z-vs-pc-site-3d} and $2.7047$ for Fig.\ \ref{fig:z-vs-ln(1-pc)-site-3d}, both very near to our predicted universal asymptotic value of $zp_{c} =  2.7351$ of Eq.\ (\ref{eq:approx3d}).

%Compare the two figures, we can see that when we want to get the slope approach the predicted value of $2.7351$, $z$ versus $1/p_{c}$ ($2.7217$) is better than $z$ versus $-1/ln(1-p_{c})$ ($2.7047$). On the other hand, when we plot $z$ versus $-1/ln(1-p_{c})$, the intercept is closer to 0, for data fitting lead to the intercept of $1.9248$ for $z$ versus $1/p_{c}$ and $0$ for $z$ versus $-1/ln(1-p_{c})$. 

\subsection{Bond percolation}
\begin{figure}[htbp] %  figure placement: here, top, bottom, or page
\centering
\includegraphics[width=3.3in]{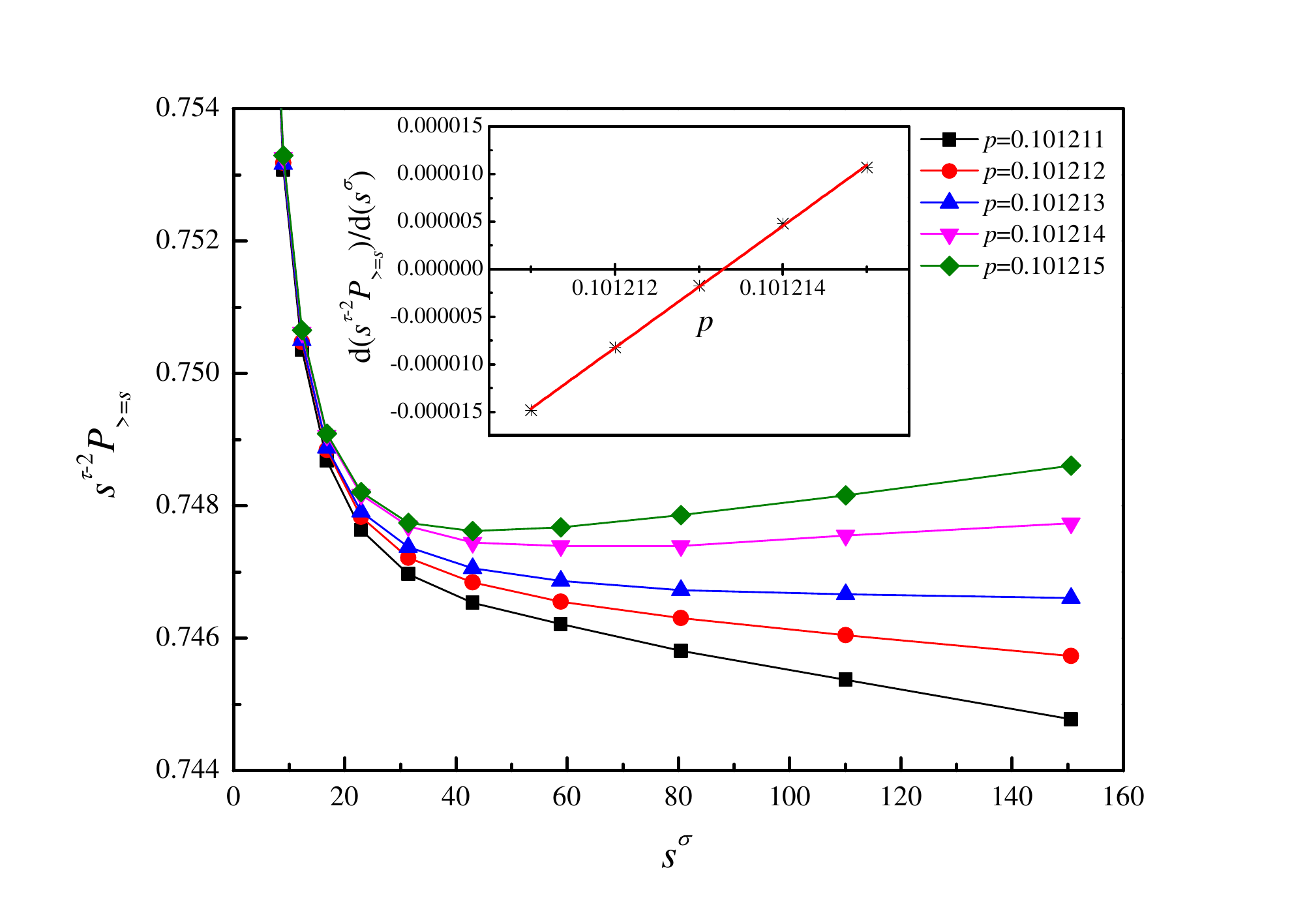}
\caption{Plot of $s^{\tau-2}P_{\geq s}$ vs\ $s^{\sigma}$ with $\tau = 2.18905$ and $\sigma = 0.4522$ for bond percolation of the \textsc{bcc}-1,2 lattice under different values of $p$. The inset indicates the slope of the linear portions of the curves shown in the main figure as a function of $p$, and the estimated value of $p_{c} = 0.1012133$ can be deduced from the $p$ intercept.}
\label{fig:bcc-1-2-sigma-bond}
\end{figure}

\begin{figure}[htbp] %  figure placement: here, top, bottom, or page
\centering
\includegraphics[width=3.3in]{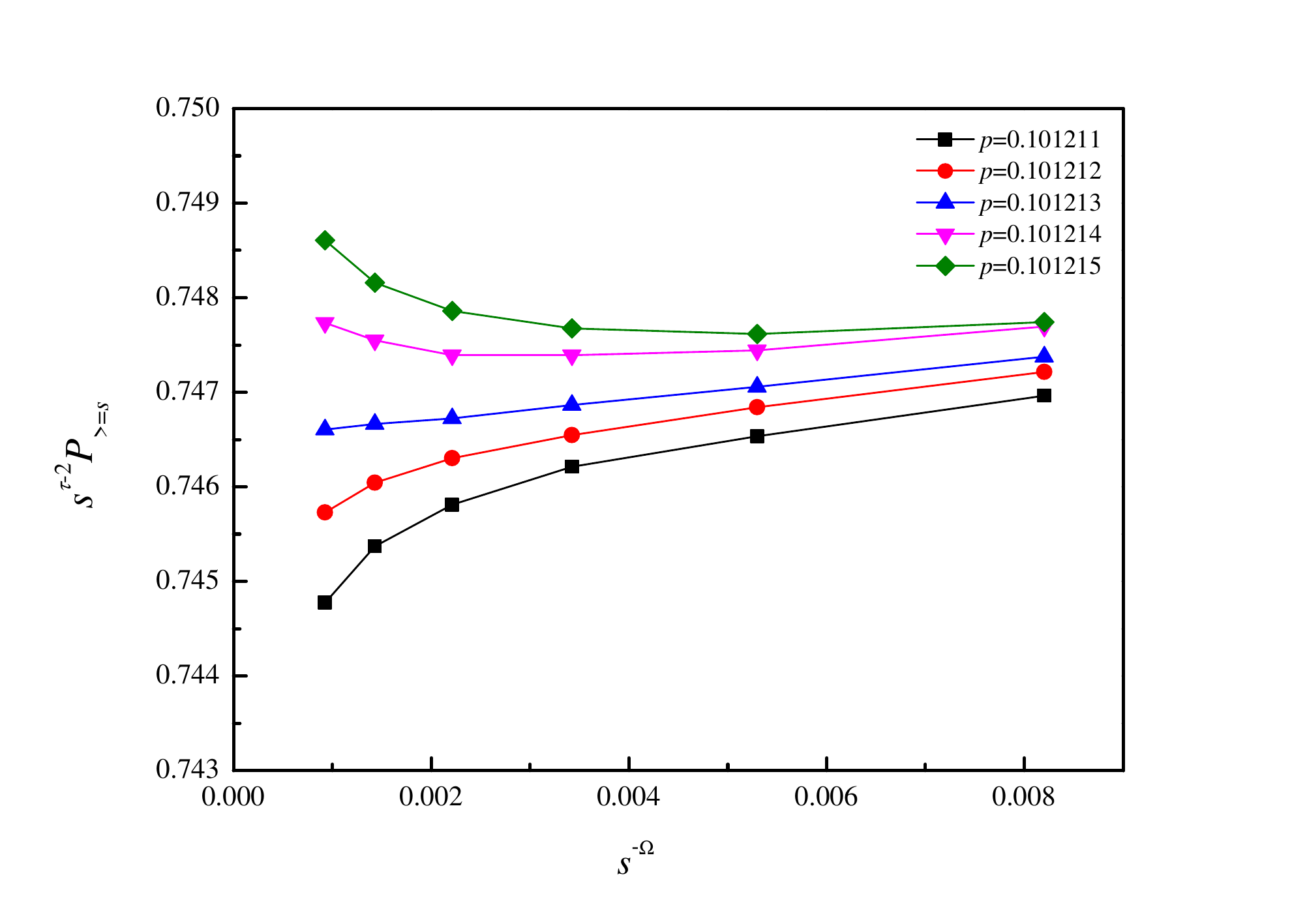}
\caption{Plot of $s^{\tau-2}P_{\geq s}$ vs $s^{-\Omega}$ with $\tau = 2.18905$ and $\Omega = 0.63$ for bond percolation of the \textsc{bcc}-1,2 lattice under different values of $p$.}
\label{fig:bcc-1-2-omega-bond}
\end{figure}

For bond percolation on \textsc{bcc}-1,2 lattice, under probabilities $p = 0.101211$, $0.101212$, $0.101213$, $0.101214$, and $0.101215$, the corresponding plots are shown in 
%relation of $s^{\tau-2}P_{\geq s}$ versus\ $s^{\sigma}$ and $s^{-\Omega}$ with $\tau = 2.18905$, $\sigma = 0.4522$, and $\Omega = 0.63$ is shown in
Figs.\ \ref{fig:bcc-1-2-sigma-bond} and \ref{fig:bcc-1-2-omega-bond}. Similar to the procedure for site percolation, the bond percolation threshold $p_{c} = 0.1012133(7)$ is estimated for this system.  Actually, it turns out that this system is identical to the \textsc{sc}-3,4 lattice, and in Ref.\ \cite{XunZiff2020} we found the identical value of the threshold for that lattice.  The simulation results for the other 12 bond lattices that we considered (including \textsc{bcc}-1,2,3, \textsc{fcc}-1,2, \textsc{fcc}-1,2,3, and \textsc{sc}-1,...,$n$ ($5 \le n \le 13$))  are shown in the Supplementary Material \cite{XunZiff2021supplementary} in Figs.\ 7-30, and the corresponding bond percolation thresholds are summarized in Table \ref{tab:bondperholds3d}.

%\setcitestyle{super,open={},close={}}
\begin{table}[htb]
\caption{Bond percolation thresholds for the \textsc{sc}, \textsc{bcc}, and \textsc{fcc} lattices with various ranges of NN. The values for \textsc{sc}-1,2, \textsc{sc}-1,2,3 and \textsc{sc}-1,2,3,4 lattices come from Ref.\ \cite{XunZiff2020b}, while the results for other lattices were determined here.  The only other previous value seems to be 0.0991(5) for \textsc{bcc}-1,2 of Ref.\ \cite{JerauldScrivenDavis1984}.}
\begin{tabular}{cccc}
\hline\hline
    lattice                     & $z$   & $p_{c}$       & $zp_{c}$  \\ \hline
    \textsc{bcc}-1,2            & 14    & 0.1012133(7)  & 1.4170  \\
    \textsc{sc}-1,2             & 18    & 0.0752326(6)  & 1.3542  \\
    \textsc{fcc}-1,2            & 18    & 0.0751589(9)  & 1.3529  \\
    \textsc{sc}-1,2,3           & 26    & 0.0497080(10) & 1.2924  \\
    \textsc{bcc}-1,2,3          & 26    & 0.0492760(10) & 1.2812  \\ 
    \textsc{sc}-1,2,3,4         & 32    & 0.0392312(8)  & 1.2554  \\
    \textsc{fcc}-1,2,3          & 42    & 0.0290193(7)  & 1.2188  \\ 
    \textsc{sc}-1,...,5         & 56    & 0.0210977(7)  & 1.1815  \\
    \textsc{sc}-1,...,6         & 80    & 0.0143950(10) & 1.1516  \\
    \textsc{sc}-1,...,7         & 92    & 0.0123632(8)  & 1.1374  \\ 
    \textsc{sc}-1,...,8         & 122   & 0.0091337(7)  & 1.1143  \\ 
    \textsc{sc}-1,...,9         & 146   & 0.0075532(8)  & 1.1028  \\
    \textsc{sc}-1,...,10        & 170   & 0.0064352(8)  & 1.0940  \\
    \textsc{sc}-1,...,11        & 178   & 0.0061312(8)  & 1.0914  \\
    \textsc{sc}-1,...,12        & 202   & 0.0053670(10) & 1.0841  \\
    \textsc{sc}-1,...,13        & 250   & 0.0042962(8)  & 1.0741  \\
\hline\hline
\end{tabular}
\label{tab:bondperholds3d}
\end{table}

\begin{figure}[htbp] %  figure placement: here, top, bottom, or page
\centering
\includegraphics[width=3.3in]{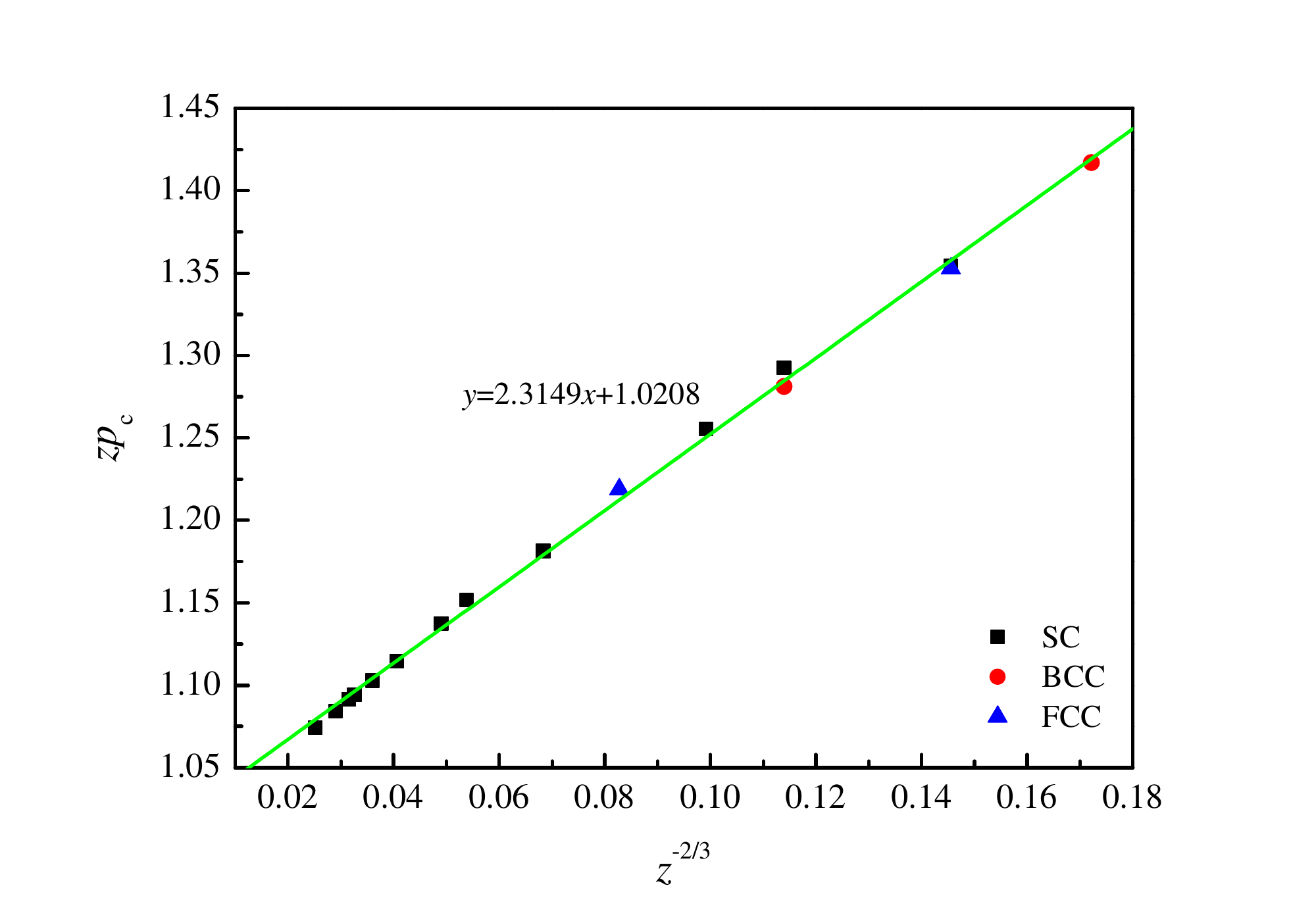}
\caption{Plot of $zp_{c}$ vs $z^{-2/3}$ for the bond percolation thresholds of the three-dimensional lattices listed in Table \ref{tab:bondperholds3d}, showing that the exponent $x=2/3$ in Eq.\ (\ref{eq:zpcbond}) gives a good representation of the finite-$z$ corrections.}
\label{fig:zpc-vs-z-bond}
\end{figure}

In Table \ref{tab:bondperholds3d}, we also list some previously known values, for lattices with various ranges of NN. We show the value of $zp_{c}$ in the fourth column of that figure. The results show that the value of $zp_{c}$ decreases from $1.4170$ to $1.0741$ with the increase of coordination number $z$ from $14$ to $250$. In Fig.\ \ref{fig:zpc-vs-z-bond}, we plot $zp_{c}$ versus $z^{-x}$, and find that a good linear fit is obtained when $x \approx 2/3$. This is in agreement with the theoretical predictions of Refs.\ \cite{XuWangHuDeng2021} and \cite{FreiPerkins2016}.  Figure \ref{fig:zpc-vs-z-bond} also exhibits an intercept very close to 1, implying that the Bethe result $p_{c} = 1/(z-1)$ accurately holds as $z \rightarrow \infty$. 

\section{Conclusions}
\label{sec:conclusions}
To summarize, in this paper, correlations between percolation threshold $p_c$ and coordination number $z$ for lattice models with compact neighborhoods, including both the asymptotic and finite-$z$ behavior, are investigated systematically. To study these correlations, extensive Monte-Carlo simulations are carried out for site and bond percolation on \textsc{bcc}, \textsc{fcc} lattices with 2nd and 3rd NN, and bond percolation on \textsc{sc} lattice with up to 13th NN. We find precise estimates of the percolation thresholds for these systems.  We also include previous results by ourselves and others to make our analysis.

For site percolation, two-dimensional Archimedean lattices and three-dimensional \textsc{sc}, \textsc{bcc}, \textsc{fcc} lattices with compact neighborhoods (up to 10th NN for Archimedean lattices, 3rd for \textsc{bcc} and \textsc{fcc} lattices, and 9th for \textsc{sc} lattice) are analyzed by plotting $z$ versus $1/p_c$ and $z$ versus $-1/\ln(1-p_c)$. We find, in a given dimension, nearly all the plots overlap in a line, with the slopes consistent with our predicted values of $zp_{c} \sim 4 \eta_c = 4.51235$ in two dimensions and $zp_{c} \sim 8 \eta_c = 2.7351$ in three dimensions.  The plot of $z$ versus $-1/\ln(1-p_c)$ gives a good fit of the behavior including the finite-$z$ corrections, with no additional adjustable parameters.

% Then the asymptotic behavior, as well as finite-$z$ corrections, of percolation on regular lattices with compact neighborhoods both in 2D and 3D are analyzed. For large $z$, we predicted their asymptotic behavior of $zp_c = 2^{d} \eta_{c}$ for site percolation, and $p_c = 1/(z-1)$ for bond percolation. The finite-$z$ effect in the simulation can be taken into account by Eqs. (\ref{eq:z}) and (\ref{eq:general}) for site percolation, and $zp_{c} - 1 \simeq a_{1}z^{-x}$ for bond percolation.

For bond percolation in three dimensions, the thresholds of \textsc{sc}, \textsc{bcc} and \textsc{fcc} lattices with compact neighborhoods (up to 3rd for \textsc{bcc} and \textsc{fcc} lattices, and 13th for \textsc{sc} lattice) confirm the finite-$z$ corrections of Eq.\ (\ref{eq:zpcbond}) with $x=2/3$ predicted by Frei and Perkins \cite{FreiPerkins2016} and Xu et al.\ \cite{XuWangHuDeng2021}, and verify that Bethe-lattice behavior $p_c \sim 1/(z-1)$ holds for large $z$.

The work in this paper indicates that the asymptotic behavior of $zp_{c}$ for compact neighborhoods, at least in two and three dimensions, is universal, depending only upon the dimension of the system but not on the type of lattice.  Of course, this universality is not as strong as the universality of critical exponents and scaling functions, which applies to all systems of a given dimensionality.  Here, the behavior depends on whether the percolation type is site or bond.  For bond percolation, the universality in the formula for $p_c$ is rather robust since for large $z$, a critical system acts like a Bethe lattice.  For site percolation, the universality we discuss here refers to all systems of compact neighbors that fall in a circular or spherical region.  If the range of the neighborhood is in a different shape, such as an elongated one, then one would use Eq.\ (\ref{eq:approx2d}) with  $\eta_c$ being the continuum threshold for objects of that shape.  Note that for both site and bond percolation, $p_c \sim c/z$ for large $z$, but the coefficient $c$ differs for the two types of percolation.  It would be interesting of course to further investigate these ideas for neighborhoods of different shapes, and in higher dimensions. (Systems of other shapes have been investigated in Refs.\  \cite{GoukerFamily83}, \cite{KozaPola2016}, and \cite{Malarz2020}, for example.) 

Overall, we see that percolation with extended-range bonds is interesting for both site and bond percolation, has many connections with literature in the field, and shows a form of universality.  The formulas of Eq.\ (\ref{eq:zpcsite}) (for site percolation) and Eq.\ (\ref{eq:zpcbond}) (for bond percolation) can be used to get good estimates for the thresholds for extended-range percolation models, as we have verified here for many systems.

\section{Acknowledgments}
The authors are grateful to the Advanced Analysis and Computation Center of CUMT for the award of CPU hours to accomplish this work. This work is supported by ``Fundamental Research Funds for the Central Universities” under Grant No.\ 2020ZDPYMS31.

\bibliographystyle{unsrt}
\bibliography{bibliography.bib}

\end{document}